\pdfoutput=1 

\documentclass[%
aip, 
pof, 
numerical,
amsmath,
amssymb,
reprint,
shortbibliography
]{revtex4-1}

\usepackage[utf8]{inputenc}    
\usepackage[T1]{fontenc}       
\usepackage[abbreviations]{glossaries-extra}
\usepackage{mathptmx}          
\usepackage{amssymb}
\usepackage{amsmath}
\usepackage{amsthm}
\usepackage{pgfmath}\pgfmathsetseed{12345}
\usepackage{bm}                
\usepackage{siunitx}           
\usepackage{CJKutf8}           
\usepackage{booktabs}          
\usepackage{tabularx}          
\usepackage{longtable}         
\usepackage{array}             
\usepackage{makecell}          
\usepackage{multirow}          
\usepackage{threeparttable}    
\usepackage{dcolumn}           
\usepackage{graphicx}          
\usepackage[export]{adjustbox} 
\usepackage{float}             
\usepackage{tikz}              
\usepackage{pgfplots}          
\usetikzlibrary{arrows.meta, positioning} 
\usepackage{hyperref}          
\usepackage{cleveref}          
\usepackage{color}             
\usepackage[dvipsnames, table]{xcolor} 
\usepackage{soul}              
\usepackage[most]{tcolorbox}   
\usepackage{lineno}
\usepackage{enumitem}          
\usepackage{everypage}         
\usepackage{etoolbox}          
\usepackage{comment}           
\usepackage{blindtext}         
\usepackage{todonotes}         

\makeatletter \@ifclassloaded{elsarticle}
{ 
    \bibliographystyle{elsarticle-num-names}
    \biboptions{sort&compress}
}{ 
    \usepackage[sort&compress]{natbib}
} \makeatother

\apptocmd{\thebibliography}{\setlength{\itemsep}{0pt}}{}{}

\usepackage{acro}
\acsetup{
  list/display = used,
  list/sort = true,
  list/template = longtable, 
  list/heading = section*,
  list/name = Nomenclature,
  list/preamble = List of acronyms used in this paper:,
}

\DeclareAcronym{HPC}{
  short = HPC ,
  long  = high performance computing 
}
\DeclareAcronym{NS}{
  short = NS ,
  long  = Navier--Stokes 
}
\DeclareAcronym{NSF}{
  short = NSF ,
  long  = Navier--Stokes--Fourier 
}
\DeclareAcronym{NSFK}{
  short = NSFK ,
  long  = Navier--Stokes--Fourier--Korteweg 
}
\DeclareAcronym{NSFE}{
  short = NSFE,
  long =  Navier--Stokes--Fourier equation,
  short-plural-form = NSFEs,
  long-plural-form = Navier--Stokes--Fourier equations
}
\DeclareAcronym{PDE}{
  short = PDE,
  long = partial differential equation,
  short-plural-form = PDEs,
  long-plural-form = partial differential equations
}
\DeclareAcronym{EoS}{
  short = EoS,
  long = equation of state,
  short-plural-form = EoS,
  long-plural-form = equations of states
}
\DeclareAcronym{w.r.t.}{
  short = w.r.t. ,
  long  = with respect to 
}
\DeclareAcronym{BE}{
  short = BE ,
  long  = Boltzmann transport equation 
}
\DeclareAcronym{BGK}{
  short = BGK ,
  long  = Bhatnagar--Gross--Krook
}
\DeclareAcronym{BGK-BE}{
  short = BGK-BE ,
  long  = Bhatnagar-Gross-Krook--Boltzmann equation 
}
\DeclareAcronym{MB}{
  short = MB ,
  long  = Maxwell--Boltzmann 
}
\DeclareAcronym{GH}{
  short = GH ,
  long  = Grad--Hermite
}
\DeclareAcronym{DF}{
  short = DF ,
  long  = distribution function
}
\DeclareAcronym{PDF}{
  short = PDF ,
  long  = probability distribution function
}
\DeclareAcronym{EDF}{
  short = EDF ,
  long  = equilibrium distribution function
}
\DeclareAcronym{QE}{
  short = QE ,
  long  = quasi-equilibrium
}
\DeclareAcronym{SE}{
  short = SE ,
  long  = shifted-equilibrium
}
\DeclareAcronym{QEDF}{
  short = QEDF ,
  long  = quasi-equilibrium distribution function
}
\DeclareAcronym{SEDF}{
  short = SEDF ,
  long  = shifted-equilibrium distribution function
}
\DeclareAcronym{DDF}{
  short = DDF ,
  long  = double-distribution function
}
\DeclareAcronym{TRT}{
  short = TRT ,
  long  = two-relaxation-time
}
\DeclareAcronym{CE}{
  short = CE ,
  long  = Chapman--Enskog
}
\DeclareAcronym{CFD}{
  short = CFD ,
  long  = computational fluid dynamics
}
\DeclareAcronym{CFL}{
short = CFL ,
long = Courant--Friedrichs--Lewy
}
\DeclareAcronym{FV}{
  short = FV ,
  long  = finite volume
}
\DeclareAcronym{FD}{
  short = FD ,
  long  = finite difference
}
\DeclareAcronym{AMR}{
  short = AMR ,
  long  = adaptive mesh refinement
}
\DeclareAcronym{AAR}{
  short = AAR ,
  long  = adaptive algorithm refinement
}
\DeclareAcronym{AMAR}{
  short = AMAR ,
  long  = adaptive mesh and algorithm refinement
}
\DeclareAcronym{SAMR}{
  short = SAMR ,
  long  = block-structured adaptive mesh refinement
}
\DeclareAcronym{LBM}{
  short = LBM ,
  long  = lattice Boltzmann method
}
\DeclareAcronym{LBE}{
  short = LBE ,
  long  = lattice Boltzmann equation
}
\DeclareAcronym{LB}{
  short = LB ,
  long  = lattice Boltzmann
}
\DeclareAcronym{DVBM}{
  short = DVBM ,
  long  = discrete velocity Boltzmann method
}
\DeclareAcronym{UGKS}{
  short = UGKS ,
  long  = unified gas kinetic scheme
}
\DeclareAcronym{DUGKS}{
  short = DUGKS ,
  long  = discrete unified gas kinetic scheme
}
\DeclareAcronym{ELBM}{
  short = ELBM ,
  long  = entropic lattice Boltzmann method
}
\DeclareAcronym{LBGK}{
  short = LBGK ,
  long  = lattice Boltzmann Bhatnagar--Gross--Krook
}
\DeclareAcronym{IC}{
  short = IC ,
  long  = initial condition
}
\DeclareAcronym{BC}{
  short = BC ,
  long  = boundary condition
}
\DeclareAcronym{MUSCL}{
  short = MUSCL ,
  long  = monotonic upstream-centered scheme for conservation Laws
}
\DeclareAcronym{KL}{
  short = KL ,
  long  = Kullback-Leibler
}
\DeclareAcronym{2D}{
  short = 2D ,
  long  = two-dimensional
}
\DeclareAcronym{1D}{
  short = 1D ,
  long  = one-dimensional
}
\DeclareAcronym{MRT}{
  short = MRT ,
  long  = multiple relaxation time
}
\DeclareAcronym{MRT-LB}{
  short = MRT-LB ,
  long  = multiple relaxation time lattice Boltzmann
}

\newcommand{\rebuttalUnmark}[1]{{\color{black}{#1}}}

\begin{document}


\title[]{
Local kinetic sensors for adaptive mesh and algorithm refinement
}
\date{\today}
\affiliation{Computational Kinetics Group, Department of Mechanical and Process Engineering, ETH Zürich, 8092 Zürich, Switzerland}
\altaffiliation{Group website: \url{https://ckg.ethz.ch/}}
\author{R.M.~Strässle}
\email{rubenst@ethz.ch}
\thanks{Corresponding author.}
\affiliation{Computational Kinetics Group, Department of Mechanical and Process Engineering, ETH Zürich, 8092 Zürich, Switzerland}
\author{S.A.~Hosseini}
\email{shosseini@ethz.ch}
\affiliation{Computational Kinetics Group, Department of Mechanical and Process Engineering, ETH Zürich, 8092 Zürich, Switzerland}
\author{I.V.~Karlin}
\email{ikarlin@ethz.ch} 
\affiliation{Computational Kinetics Group, Department of Mechanical and Process Engineering, ETH Zürich, 8092 Zürich, Switzerland}


\begin{abstract} 

    This paper presents novel refinement sensors for the application to adaptive mesh and algorithm refinement (AMAR) with kinetic models, such as discrete velocity and lattice Boltzmann methods.
    While refinement criteria for AMAR based on macroscopic variables can be replicated in a purely local, and therefore more scalable, way, the main advantage that can be leveraged when working with discrete velocity and lattice Boltzmann methods is the accessibility of information from the one-particle distribution function.
    With this accessibility, a novel palette of refinement sensors is introduced, allowing for a set of neatly tailored refinement criteria applicable to resolve characteristic flows features in many relevant domains of fluid mechanics, for instance, those emerging in compressible, turbulent, and non-equilibrium flows or non-ideal fluids.
    After detailed validation, novel refinement sensors are showcased for the application of adaptive mesh refinement (AMR) to a discrete velocity Boltzmann solver for compressible, viscous, and non-equilibrium flows, demonstrating promising results.
    The proposed sensors establish an accurate, efficient and scalable approach to kinetic simulations with AMAR, offering a valuable tool for studying complex problems in fluid dynamics and paving the way for future extensions to more specific flow problems.
    
    \vspace{0.5cm}
    
    
\end{abstract}
\maketitle


\section{Introduction\label{sec:Introduction}}

Detailed understanding of fluid flows is of paramount importance for science and engineering. 
The development of reliable, accurate, and efficient numerical methods for the simulation of fluid dynamics, especially on large-scale clusters, has been a topic of intense research over the past few decades~\cite{pirozzoli2011numerical, yee2018recent}. 
Although different discrete approximations such as \acp{FV} and \acp{FD} to the \ac{NSF} equations were the main drivers of research in \ac{CFD}, the advent of \ac{LBM} in the late 1980s opened the door for a new class of numerical methods such as \acp{DVBM} rooted in the kinetic theory of gases~\cite{LBMBookKrueger, succiBook}. 

Besides extensions in non-classical physics, e.g., plasma dynamics on the basis of the Vlasov equation~\cite{vlasov1961many}, post-Newtonian physics such as quantum mechanics~\cite{Succi_2038}, relativistic and quantum-relativistic fluid dynamics~\cite{Succi_35yearsdown, Romatschke_Relativistic}, 
kinetic theory found its main application in the context of classical fluid mechanics, where a discrete velocity version of the Boltzmann transport equation~\cite{Boltzmann1872weitere}, which evolves a probability density function of the particle velocity in space and time, is considered.
The dynamics of observable properties of a fluid described by the Euler or \ac{NSF} equations are recovered in the hydrodynamic limit~\cite{Chapman}.
In addition to allowing for the possibility to include physics beyond \ac{NSF}, kinetic models in combination with collision models such as the \ac{BGK} approximation~\cite{bhatnagar1954model} have been shown to provide an efficient alternative to classical solvers~\cite{QianLBM, ChenLBM}, which allowed considerable advances in the past decades concerning simulation of 
non-equilibrium~\cite{Nagnibeda2009NEQ, Xu_2008_hypersonicNEQ, Wang_2014_UGKS_multicomponentNEQ, Su_NEQhardsphere}, 
turbulent~\cite{Humières_MRT, Geier_Cascaded, Geier_Cumulant, malaspinas_recursivereg, Karlin_entropicEQ, Karlin_Gibbs, bosch2015entropic, Ali_reviewEntropic, FRAPOLLI2018ELBforthermal, Teixeira1998TurbulenceModelsInLB, Jahanshaloo2013ReviewLBMturbulent}, 
compressible~\cite{FENG_FV_DBM, UGKS_14, DUGKS_compressiblecase, xu2018discrete, guo2021progress, GuoDUGKS23, hosseini2020compressibility, FengCorrection, Saadat2019, saadat2021extended, BardowMultispeedSSLBM, Wilde_SSLBM_CF, strässle2025consistent-compressible}, 
high-speed~\cite{Ali_shiftedStencils, PonD18, Bhaduria23, Kallikounis22, Kallikounis23, Ji24}, 
rarefied~\cite{mieussens2000discrete, Buet_rarefied, Li_Rarefied, YANG_Rarefied}, 
non-ideal and multiphase~\cite{Shan1993Lattice, Shan1994Simulation, Swift1995Lattice, HeDollen_2002_foundationsMultiphase, Sbragaglia2007Generalized, Sbragaglia2011Consistent, Mazloomi2015Entropic, Reyhanian20, Hosseini2022Towards, hosseini_2025_compressiblenonideal}, 
multicomponent/multi-species~\cite{Wang_2014_UGKS_multicomponentNEQ, Briant_2016_multispecies, Haack2017MultispeciesBGK, sawant_2021_consistentMulticomponent} 
and reactive flows~\cite{Hosseini2018MultispeciesReactive, sawant_2021_AlatticeReactive, sawant_2022_consistentReactive, Sawant_2025_meanfield}, 
including modeling of equilibrium and non-equilibrium thermodynamic effects on the basis of 
first principles~\cite{Hosseini2022Towards, sawant_2021_consistentMulticomponent, sawant_2022_consistentReactive}. 
In parallel, further appropriations from the world of computational physics and conventional \ac{CFD} have also been implemented and developed.

One of such commonly used computational techniques is \ac{AMAR}.
Both \ac{AAR} and \ac{AMR} are designed to enhance the ratio between computational efficiency and physical accuracy of simulations on a computational grid.
In the case of \ac{AAR} this is accomplished by dynamically adjusting the solution algorithm or physical modeling approach, whereas, the resolution is dynamically adjusted in the case of \ac{AMR}.
Both techniques can lead to reduced computational cost and memory usage in regions where less physical accuracy or coarser resolution is sufficient by adaptively refining the algorithm or computational grid in regions of interest. 
These refinement techniques are beneficial for a variety of different scenarios, including flows containing localized regions of high complexity such as
(1) highly off-equilibrium dominated regions,
(2) boundary layers around objects, vortex dominated, transitional or turbulent flows,  
as well as sharp gradients of field variables such as those exhibited in 
(3) compressible and high-speed flows, due to the presence of contact discontinuities, shocks and rarefaction waves,
(4) rarefied gases  and
(5) non-ideal and multiphase flows, due to interfaces and phase changes,
as well as interaction and reaction zones such as those found in 
(6) multicomponent/multi-species 
(7) and reactive flows.
Besides the challenges associated with the incorporation of \ac{AMAR} in an existing solution methodology, such as strict conservation and leveraging the full potential of parallel processing on modern distributed memory machines, it becomes evident from the above illustrated application examples, that the local refinement indicators of the flow, so-called refinement sensors, play a crucial role to the success of \ac{AMAR} techniques.

In connection with kinetic solvers, the incorporation of \ac{AMR} has been demonstrated, e.g., in \cite{DorschnerRefinementELBM, Guzik, EitelAmor_AMR_LBM, He_AMR_LBM, Schukmann_AMR_LBM, Fakhari_FD_LBM, Huang_AMR_RLBFS},
\rebuttalUnmark{and also combined \ac{AMAR} has been demonstrated, e.g., in \cite{Kolobov_2007_Unified}.}
A few \rebuttalUnmark{kinetic} applications can also be considered \rebuttalUnmark{pure} \ac{AAR} \cite{Kallikounis_multiscale, 2016Entropic, PonD18, Coreixas2020adaptivevelocity}; 
for example, shifted lattice models and the particles on demand method \cite{PonD18, Coreixas2020adaptivevelocity}, since they adapt reference frames based on the local or global fluid velocity and temperature, 
or adaptive quadrature orders of the discrete velocity stencils \cite{Kallikounis_multiscale}.
Another type of kinetic model that could be considered \ac{AAR} is the class of entropic lattice Boltzmann methods, which employ adaptive relaxation paths to the collision process based on entropy considerations, with some of them introducing additional dissipation to ensure stability.
The adaptive relaxation is based on the equal entropy ($H$-function) condition of the mirror state in the case of the entropic lattice Boltzmann method (ELBM) \cite{2016Entropic, Ali_reviewEntropic}, and on  
the difference between microscopic and macroscopic entropy ($H$-function) in the case of Ehrenfest's regularization steps \cite{Brownlee2006Stabilization, Brownlee2007Stability, Brownlee2008Nonequilibrium}.

Although the criticality of refinement sensors led to many contributions in the conventional CFD literature in the past, 
most notably seen with the family of sensors related to the $Q$-criterion used in strain-vortex dominated and turbulent flows, 
developments of sensors in connection with kinetic models received less attention, even though the accessibility of additional information from the distribution function appears to be a considerable advantage.
The kinetic models employing \ac{AMR} mostly relied on macroscopic sensors, and the \ac{AAR} examples can be considered specific sub-cases of kinetic models.
The exceptions known to the authors which can be labeled kinetic refinement sensors are given by a local estimate of the lattice Knudsen number, as introduced in the \ac{AMR} context in \cite{Thorimbert_KNsensor} and further demonstrated in the \ac{AAR} context in \cite{Coreixas2020adaptivevelocity}, as well as the entropic estimate $\alpha$ or the difference between microscopic and macroscopic entropy, which have been purely used in the context of \ac{AAR} with the class of entropic lattice Boltzmann methods mentioned above \cite{Ali_reviewEntropic, Brownlee2006Stabilization}.
More generic refinement sensors, which permit broader applicability to \ac{AMAR} of kinetic models, have been largely lacking.

For this reason, the present paper introduces local kinetic refinement sensors with generic applicability to adaptive mesh and algorithm refinement of kinetic models. 
The advantages of these sensors are two fold; 
On the one hand, sensors can be constructed with information which is not accessible from the macroscopic description of fluids. 
On the other hand, the information that can be accessed from macroscopic variables can also be computed from the kinetic description, with the difference that sensors based on gradients of macroscopic variables, for example, stress tensors, heat fluxes, etc., can be computed in a purely local manner as a summation over the discrete populations instead of applying, e.g., finite differences to compute gradients, leading to efficiency and scalability gains when considering parallel codes for \ac{HPC}.
Although the sensors are presented for a discrete kinetic model for compressible flows with variable Prandtl number and demonstrated for the application with \ac{AMR}, both based on our recent publications \cite{strässle2025consistent-compressible, strässle2025a-fully-conservative}, these sensors can be used for a wide range of problems and flow scenarios, and the ideas can be extended to construct more beneficial sensors for more specific flow scenarios, with some examples provided herein.

The outline of the paper is as follows:
In order to provide sufficient context for self-consistency, all necessary background related to the target hydrodynamic equations, the kinetic model \cite{strässle2025consistent-compressible}, and the mesh refinement methodology \cite{strässle2025a-fully-conservative} are provided in Section~\ref{sec:Background}.
The local kinetic refinement sensors are described in Section~\ref{sec:Refinement-sensors}, followed by a detailed evaluation in Section~\ref{sec:Validation}.
The sensors are then demonstrated with numerical examples in Section~\ref{sec:ApllicationAMR}.
Finally, a summary and conclusions are provided in Section~\ref{sec:Conslusions}.

\section{Background\label{sec:Background}}

\subsection{Kinetic model\label{sec:Kinetic-model}}

In this paper, the consistent kinetic modeling approach for compressible flows with variable Prandtl number is employed to showcase the refinement sensors.
Only the necessary concepts shall be highlighted hereafter.
For more detail, the reader is referred to \cite{strässle2025consistent-compressible}.

\subsubsection{Model description}

The kinetic model is based on the \ac{BE} where the full Boltzmann collision integral~\cite{Boltzmann1872weitere} is modeled with the \ac{BGK} approximation~\cite{bhatnagar1954model}.
The resulting \ac{BGK-BE}, however, is not sufficient to capture compressible flows with variable Prandtl number on the Navier--Stokes--Fourier level in the hydrodynamic limit, as there are two well-known shortcomings that need to be addressed: 
\begin{itemize}
    \item[(1)] Variable specific heat capacity for polyatomic molecules: The distribution function has to be extended in order to account for the internal roto-vibrational degrees of freedom of the gas molecules in a polyatomic gas.
    \item[(2)] Variable Prandtl number: The \ac{BGK} collision operator results in the restriction of a Prandtl number of unity.
\end{itemize}
To resolve the first issue, the idea of Rykov~\cite{rykov_model_1976} is followed, who extended the BGK framework to polyatomic gases by introducing additional relaxation mechanisms to account for the additional degrees of freedom. For computational convenience, this idea was later reformulated, mainly based on the lattice Boltzmann picture, by introducing a second distribution function $g$, evolving according to another Boltzmann transport equation, which explicitly represents the evolution of internal roto-vibrational energy modes of the gas. 
This description further allowed to extend the second distribution to carry some, or the full, part of the total energy, as the $g$- and reduced $f$-distribution can be linked to constitute the necessary information~\cite{ProbingDoubleDist2024}. 
An option to resolve the second limitation is given by the application of the generalized-BGK collision operator, guided by the \ac{QE} approximation approach~\cite{gorban1994, gorban1991quasi, Gorban2006Quasi}. 
The latter approach decomposes the dynamics of the system into fast and slow modes, which is very practical for systems possessing different time scales of physical processes, such as a Prandtl number of non-unity,
\begin{equation}
  {\rm Pr} = \frac{C_p \mu}{\kappa} = \frac{\nu}{\alpha}
  ,
\end{equation}
which expresses the ratio of momentum and thermal diffusion. 
In the case where the heat flux is regarded as the "slow" variable, the thermal conductivity $\kappa$ (or thermal diffusivity $\alpha$) can be related to the "slow" relaxation time, while the dynamic (shear) viscosity $\mu$ (or kinematic viscosity $\nu$) is proportional to the "fast" relaxation time, and vice versa.

In summary, to account for both issues, the double-distribution quasi-equilibrium formalism, i.e. a combination of two of the most widely used approaches, is applied.
The transport equations become 
\begin{equation}
    \partial_t \{ f, g\} (\bm{v},\bm{r},t)
    + \bm{v} \cdot \bm{\nabla} \{ f, g\}  (\bm{v},\bm{r},t)
    = \Omega_{\{f,g\}}
,\label{new_stream_f_serial}
\end{equation}
where the collision operator is written as
\begin{multline}
\Omega_{\{f,g\}}
    = - \frac{1}{\tau_1} \left[ \{ f, g\}  (\bm{v},\bm{r},t) - \{ f^*, g^*\}  (\bm{v},\bm{r},t) \right] 
    \\
    - \frac{1}{\tau_2} \left[ \{ f^*, g^*\}  (\bm{v},\bm{r},t) - \{ f^{\rm eq}, g^{\rm eq} \}  (\bm{v},\bm{r},t) \right]
.\label{new_collision_f_serial}
\end{multline}
Here, the particle velocity is designated by $\bm{v}$ while $\bm{r}$ marks the position in space and $t$ the time. 
The \acp{PDF} and the local \acp{EDF} are represented by $f \left(\bm{r},\bm{v},t\right)$, $g \left(\bm{r},\bm{v},t\right)$, and $f^{\rm eq} \left(\bm{r},\bm{v},t\right)$, $g^{\rm eq} \left(\bm{r},\bm{v},t\right)$, respectively.
Furthermore, $f^* \left(\bm{r},\bm{v},t\right)$ and $g^* \left(\bm{r},\bm{v},t\right)$ are the \acp{QEDF}, and
the parameters $\tau_1$ and $\tau_2$ are the relaxation times that control the relaxation rate of the distribution functions towards the quasi-equilibria and the equilibria attractors.
For a strict relaxation time hierarchy
\begin{equation}
    \tau_1\le \tau_2,
\end{equation}
the resulting \ac{QE} kinetic model fulfills Boltzmann’s $H$-theorem, which states that, for solutions of the Boltzmann equation, the production of the $H$-function,
\begin{equation}
    H(\{f,g\}) = \int \{f,g\} \ln \left( \{f,g\} \right) d \bm{v}
    ,\label{eq:H-function}
\end{equation}
due to the collision integral is non-positive definite, i.e.,
\begin{equation}
    \sigma_H(\{f,g\}) = \int  \Omega_{\{f,g\}} \ln \{f,g\} \mathrm{d}\bm{v} \le 0
    ,\label{eq:Htheorem}
\end{equation}
expressing irreversible entropy production and a monotonic decay of $H$ toward its minimum.
At its minimum, the $H$-function is directly related to the thermodynamic entropy $s$ of a gas in the single-distribution setting through
\begin{equation}
    s = -k_B H(f^{\rm eq})
    ,\label{eq:relation-H-entropy}
\end{equation}
where $k_B$ denotes the Boltzmann constant.
The $H$-function may therefore be interpreted as a non-equilibrium extension of the thermodynamic entropy when the distribution function departs from the 
equilibrium distribution family, and the monotonic decrease of $H$ implied by the $H$-theorem corresponds to non-negative entropy production in accordance with the second law of thermodynamics.

As for models without variable specific heat capacity and variable Prandtl number, the local equilibrium of the reduced $f$-distribution is given by the \ac{MB} equilibrium distribution function, parametrized by mass density $\rho$, velocity $\bm{u}$ and temperature $T$, as
\begin{equation}
    f^{\rm eq} (\bm{v},\bm{r},t)
    = \frac{n }{(2\pi R T)^{D/2}} \exp{\left[-\frac{{{|}\bm{v}-\bm{u}{|}}^2}{2RT}\right]}
    ,\label{eq:MB-equilibrium}
\end{equation}
where $n=\rho / m$ is the particle number density with the particle mass $m$ and mass density $\rho$, $D$ is the dimension of the physical space and $R$ designates the specific gas constant ($m=1$ and $R=1$ are used for simplicity in this manuscript).
The \ac{MB} \ac{EDF} annuls the Boltzmann collision integral, and is also the minimizer of the $H$-function,
under constraints of locally conserved moments, density, momentum and total energy, i.e., 
\begin{equation}
    f^{\rm eq} = \mathrm{argmin} \ H (f) \Big|_{\rho, \rho \bm{u}, E}
    ,\label{eq:minH_f_DDF}
\end{equation}
where conserved density and momentum are found as, 
\begin{gather}
    \rho 
    = \int f^{\rm eq} d\bm{v}
    = \int f^{*} d\bm{v} 
    = \int f d\bm{v}     
    ,\label{eq_con_fstar1}
\\
    \rho \bm{u} 
    = \int \bm{v} f^{\rm eq} d\bm{v}
    = \int \bm{v}  f^{*} d\bm{v} 
    = \int \bm{v} f d\bm{v}
    ,\label{eq_con_fstar2}
\end{gather}
and the total energy represents the sum of the internal energy $U$ and the kinetic energy $K$ as
\begin{equation}
    E = U + K = \rho C_v T + \rho \frac{u^2}{2}
    ,\label{eq:Total_energy_MB}
\end{equation}
for a perfect gas.

The reduced $f^{\rm eq}$ becomes dependent on the second distribution function $g$, which carries parts of the additional energy due to the presence of non-translational degrees of freedom.
While different partitions of energy on the $g$-distribution are possible, this work was restricted to the two most popular partitions, i.e, the forms where either the total energy or the internal non-translational energy are put on $g$, respectively. 
The total energy is therefore defined as 
\begin{equation}
    E 
    = \int g^{\rm eq} d\bm{v} 
    = \int g^{*} d\bm{v} 
    = \int  g d\bm{v} 
    ,\label{eq_con_gstar_total}
\end{equation}
for the total energy split and, using the notation $v^2 = \bm{v}\cdot\bm{v}$, as 
\begin{multline}
    E 
    = \int g^{\rm eq} + \frac{v^2}{2} f^{\rm eq} d\bm{v} 
    = \int g^{*} + \frac{v^2}{2} f^{*}  d\bm{v} 
    = \int g + \frac{v^2}{2} f d\bm{v}
    ,\label{eq_con_gstar_rykov}
\end{multline}
for the internal non-translational energy split.
Consequently, the equilibrium of the second distribution is defined as a reparameterization of $f^{\rm eq}$, as
\begin{equation}
    g^{\rm eq} 
    = \left( {\frac{U^{\rm ntr}}{\rho}} + \frac{v^2}{2} \right) f^{\rm eq}
    = \left (C_vT-\frac{RDT}{2} + \frac{v^2}{2} \right) f^{\rm eq}
    ,\label{energy_eq_1_total}
\end{equation}
for the total energy split and as 
\begin{equation}
    g^{\rm eq} 
    = {\frac{U^{\rm ntr}}{\rho}} f^{\rm eq}
    = \left (C_v T - \frac{RDT}{2} \right) f^{\rm eq}
    ,
    \label{energy_eq_1_rykov}
\end{equation}
for the internal non-translational energy split, respectively, where
\begin{equation}
    U^{\rm ntr} = U - U^{\rm tr} = U - \rho \frac{DR}{2}
\end{equation}
represents the internal energy associated with the non-translational degrees of freedom.

Furthermore, the \acp{QEDF} are defined as the minimizer of the $H$-function subject to the locally conserved density, momentum and total energy, as well as additional quasi-conserved slow fields.
In the context of applying the \ac{QE} notion to capture variable Prandtl numbers, the pressure tensor and the heat flux vector mark the quasi-conserved fields, with altering conditions depending on $\{ \mathrm{Pr}\leq1$, $\mathrm{Pr}\geq1 \}$.
In practical applications, only approximate expressions for the \acp{QEDF} are usually available, such as Grad's moment approximations \cite{grad1949kinetic} or \rebuttalUnmark{through} the triangle entropy method \cite{gorban1991quasi}.

\subsubsection{Hydrodynamic limit and the Navier--Stokes--Fourier equations} 

The conclusions of the multiscale analysis in the form of the Chapman-Enskog expansion~\cite{Chapman} shall be highlighted at this point.
In the hydrodynamic limit, the specified system recovers the \ac{NSF} equations, i.e.,
\begin{gather}
    \partial_t\rho + \bm{\nabla} \cdot \rho \bm{u} 
    = 0
    ,
    \\
    \partial_t (\rho \bm{u}) + \bm{\nabla} \cdot \left( \rho \bm{u} \otimes \bm{u} + p\bm{I} {+}\bm{\tau}_{\rm NS} \right) 
    = 0
    ,
    \\
    \partial_t E + \bm{\nabla} \cdot \bigl[ (E + p)\bm{u} + \bm{q}_{\rm NSF}
    \bigr]
    = 0
    ,
\end{gather}
where the dissipative mechanisms are in the form of the Navier--Stokes stress tensor
\begin{multline}
    \bm{\tau}_{\rm NS} = 
     {-}\mu\left[ \bm{\nabla}\bm{u} + \bm{\nabla}\bm{u}^{\dagger}- \frac{R}{C_v}(\bm{\nabla}\cdot\bm{u})\bm{I} \right]
    \\ 
    =  {-}\mu \left[\bm{\nabla}\bm{u} + \bm{\nabla}\bm{u}^\dagger - \frac{2}{D}(\bm{\nabla}\cdot\bm{u})\bm{I} \right] 
     {-} \eta (\bm{\nabla}\cdot\bm{u})\bm{I}
    ,\label{eq:Tns}
\end{multline}
the Navier--Stokes--Fourier energy flux vector
\begin{equation}
    \bm{q}_{\rm NSF} = \bm{q}_{\rm F}+\bm{q}_{\rm H}
    ,\label{eq:NSFenergyflux}
\end{equation}
where the Fourier heat flux is
\begin{equation}
    \bm{q}_{\rm F} = - \kappa \bm{\nabla}T
    ,\label{eq:Fourier}
\end{equation}
and the viscous heating vector is
\begin{multline}
    \bm{q}_{\rm H} = \bm{u} \cdot \bm{\tau}_{\rm NS} =
    {-} \mu \Bigl[ \bm{u}\cdot\bm{\nabla}\bm{u} + \bm{u}\cdot\bm{\nabla}\bm{u}^{\dagger} - \frac{R}{C_v} {\bm{u}(\bm{\nabla}\cdot\bm{u})}\Bigr]
    \\ 
    = {-}\mu \left[\bm{u}\cdot\bm{\nabla}\bm{u} + \bm{u}\cdot\bm{\nabla}\bm{u}^\dagger - \frac{2}{D} {\bm{u}(\bm{\nabla}\cdot\bm{u})} \right] 
     {-\eta \bm{u}(\bm{\nabla}\cdot\bm{u})}
    .\label{eq:viscHeating}
\end{multline}
From the multiscale analysis it is further found that the relaxation parameters $\tau_1$ and $\tau_2$ are related to the kinematic or dynamic shear viscosity $\nu$ or $\mu$, kinematic or dynamic bulk viscosity $\zeta$ or $\eta$, and thermal diffusivity $\alpha$ or conductivity $\kappa$ as, 
\begin{gather}
    \label{visc_shear_1}
    \nu = \frac{\mu}{\rho} = \{ \tau_1, \tau_2 \}    R T
    , 
\\
    \label{visc_bluk_1}
    \zeta = \frac{\eta}{\rho} = \left(\frac{2}{D}-\frac{R}{C_v}\right) \{ \tau_1, \tau_2 \} R T
    ,
\\
    \label{visc_thermal_1}
    \alpha = \frac{\kappa}{C_p\rho} = \{ \tau_2, \tau_1 \} R T
    ,
\end{gather}
for 
\begin{equation}
    \mathrm{Pr} = \{ \leq 1, \geq 1\}
    ,
\end{equation} 
while the Prandtl number and the relaxation parameters are related through 
\begin{equation}
    \mathrm{Pr} = \left\{ \frac{\tau_1}{\tau_2}, \frac{\tau_2}{\tau_1} \right\}
    .
\end{equation}

Note that Eqs. \ref{eq:Tns} and \ref{eq:viscHeating} were written in two corresponding forms.
Multiple forms are possible, e.g., for the stress tensor,
\begin{multline}
    \bm{\tau}_{\rm NS} 
    = -2\mu\bm{E} - \lambda{\rm tr}(\bm{E})\bm{I}
    = -2\mu\bm{E} - \lambda\bm{C}
    \\
    = -2\mu\left(\bm{E}-\frac{\bm{C}}{D}\right) - \eta\bm{C}
    = -\mu\bm{S} - \eta\bm{C}
    ,\label{eq:tauNS-1} 
\end{multline}
which allows to introduce some physically meaningful tensors, using the following notation.
After identifying the second viscosity as
\begin{equation}
    \lambda = - \frac{\mu R}{C_v}
    ,
\end{equation}
and the dynamic bulk viscosity as
\begin{equation}
    \eta = \frac{2\mu}{D} + \lambda = \mu \left( \frac{2}{D} - \frac{R}{C_v} \right)
    ,
\end{equation}
the symmetric part of $\bm{\nabla}\bm{u}$ is the rate-of-strain tensor  
\begin{equation}
    \bm{E} = \frac{1}{2}\bm{\nabla}\bm{u} +  \frac{1}{2}\bm{\nabla}\bm{u}^{\dagger} 
    .\label{eq:rate-of-strain}
\end{equation}
The trace of the rate-of-strain tensor is the divergence of the velocity field, ${\rm tr}(\bm{E}) = \bm{\nabla}\cdot\bm{u}$, resulting in the rate-of-compression tensor 
 \begin{equation}
     \bm{C} = (\bm{\nabla}\cdot\bm{u}) \bm{I}
     .\label{eq:rate-of-compression}
 \end{equation}
The rate-of-shear tensor,
\begin{equation}
    \bm{S} = \bm{\nabla}\bm{u} + \bm{\nabla}\bm{u}^{\dagger} 
    - \frac{2}{D} (\bm{\nabla}\cdot\bm{u}) \bm{I}
    ,\label{eq:rate-of-shear}
\end{equation}
emerges after separation of all trace-free parts. 
Similar considerations can be done for the energy balance equation.
Furthermore, an expression for the characteristic speed with which information travels in a fluid, i.e. the speed of sound $c_s$, can be defined by taking the isentropic relations for a perfect gas into account, as
\begin{equation}
\label{eq:NSspeedofsound}
    c_s = \sqrt{\frac{\partial p}{\partial \rho}\Big|}_{\delta s=0} = \sqrt{\gamma \frac{p}{\rho}}
    .
\end{equation}

\subsubsection{Discrete kinetic model} 

Discretization of the distribution functions $\{f,g\} \rightarrow \{f_i,g_i\} \ \forall \ i$ , where $\{f_i,g_i\}$ are referred to as populations, is operated via expansion using the Hermite orthonormal polynomials and application of the Gauss-Hermite quadrature with a set of $Q$ discrete velocities $\bm{v}_i$, where $i \in \{0, Q-1\}$. 
The phase-space-discretized system of hyperbolic \acp{PDE} then reads
\begin{equation}
     \partial_t \{ f_i, g_i\} (\bm{r},t)
    + \bm{v}_i \cdot \bm{\nabla} \{ f_i, g_i\}  (\bm{r},t)
    = \Omega_{\{f_i,g_i\}}
    ,
\end{equation}
with the discrete collision operator
\begin{multline}
    \Omega_{\{f_i,g_i\}} =  -\frac{1}{\tau_1} \left[ \{ f_i, g_i\} (\bm{r},t)  - \{ f_i^*, g^*_i\} (\bm{r},t) \right] 
\\
    - \frac{1}{\tau_2} \left[ \{ f_i^*, g^*_i\}  (\bm{r},t)
    -  \{ f_i^{\rm eq}, g^{\rm eq}_i\}  (\bm{r},t) \right]
    .\label{new_collision_fi}
\end{multline}
Moments of distribution functions are computed as numerical quadratures, e.g. the conserved density, momentum and total energy as
\begin{gather}
    \label{eq_quadratur1}
    \rho 
    = \sum_{i=0}^{Q-1} f^{\rm eq}_i 
    = \sum_{i=0}^{Q-1} f^{*}_i
    = \sum_{i=0}^{Q-1} f_i 
    , 
\\
    \label{eq_quadratur2}
    \rho\bm{u} 
    = \sum_{i=0}^{Q-1} \bm{v}_i f^{\rm eq}_i 
    = \sum_{i=0}^{Q-1} \bm{v}_i f^{*}_i
    = \sum_{i=0}^{Q-1} \bm{v}_i f_i 
    ,  
\end{gather}
and  
\begin{equation}
        E 
        = \sum_{i=0}^{Q-1}  g^{\rm eq}_i 
        = \sum_{i=0}^{Q-1} g^{*}_i
        = \sum_{i=0}^{Q-1}  g_i 
        ,\label{eq_quadratur3_total}
\end{equation}
for the total energy split or
\begin{equation}
        E 
        = \sum_{i=0}^{Q-1} g^{\rm eq}_i + \frac{v_i^2}{2} f^{\rm eq}_i 
        = \sum_{i=0}^{Q-1} g^*_i + \frac{v_i^2}{2}  f^*_i
        = \sum_{i=0}^{Q-1} g_i + \frac{v_i^2}{2}  f_i 
        ,\label{eq_quadratur3_intnontrans}
\end{equation}
for the internal non-translational energy split.
The $H$-function in the discrete version becomes, cf. \cite{Karlin_1999_perfectEntropy, Ansumali_2003_MinimalEntropic},
\begin{equation}
    H(\{f,g\}) = \sum_{i=0}^{Q-1} \{f_i,g_i\} \ln \left( \frac{\{f_i,g_i\}}{w_i} \right) 
    ,\label{eq:H-function-discrete}
\end{equation}
and the discrete production of the $H$-function reads
\begin{equation}
    \sigma_H (\{f,g\}) = \sum_{i=0}^{Q-1} \Omega_{\{f_i,g_i\}} \ln \left( \frac{\{f_i,g_i\}}{w_i} \right) \leq 0
    ,\label{eq:sigma-function-discrete}
\end{equation}
where $\Omega_{\{f_i,g_i\}}$ is the discrete version of the collision operator, i.e., Eq. \ref{new_collision_fi}, and $w_i$ are the weights associated to the discrete velocities.

To capture the fundamental flow physical properties of the \ac{NSF} equations in the hydrodynamic limit, for the $f$-distribution, equilibrium moments up to order $N=3$ have to be properly recovered for the total energy split, while, for the non-translational energy split, fourth-order ($N=4$) equilibrium moments have to be properly recovered as well. 
For the $g$-distribution, regardless of the split, equilibrium moments up to order $N=2$ need to be properly recovered.
These orders of expansion require higher-order velocity sets, cf.~\cite{FrapolliMultispeed} or standard nearest-neighbor velocity sets with correction terms to the otherwise incorrectly recovered higher-order moments~\cite{saadat2021extended, Prasianakis2007}. 
In this work, all equilibria and quasi-equilibria populations were constructed with the Grad--Hermite expansion with moments up to order $N$, as
\begin{equation}
    \{ f^{\rm eq}_{i}, g^{\rm eq}_{i}, f^{*}_{i}, g^{*}_{i} \}
     = w_i \sum_{n=0}^{N} \frac{1}{n!(RT_{ref})^n} 
     \bm{a}_{n}^{\{ {\rm eq}, *\} }(\{ f, g \}) : \bm{\mathcal{H}}_{n}(\bm{v}_i)
     ,
\end{equation}
where $T_{ref}$ is reference temperature associated to the applied discrete velocities and $\bm{\mathcal{H}}_{n}$ are the Hermite-polynomial tensors.
The equilibrium moments are fed into the coefficient tensors $\bm{a}_{n}^{\rm eq}$ in order to construct the equilibria populations.
For the quasi-equilibria populations, a different set of moments is accounted for in $\bm{a}_{n}^{\rm *}$; these generally consist of equilibrium moments (to satisfy the same constraints of conserved moments in the definition of the \acp{QEDF} as minimizers of discrete $H$-function), with some additional moments stemming from the constraints for the quasi-conserved fields.
The solution via higher-order velocity sets was considered herein, where the minimal Hermite-based higher-order velocity sets (in two dimensions), i.e., the D2Q16 and the D2Q25 were used for the total and internal non-translational energy splits, respectively.

\begin{figure*}
\centering
\includegraphics[width=0.99\linewidth]{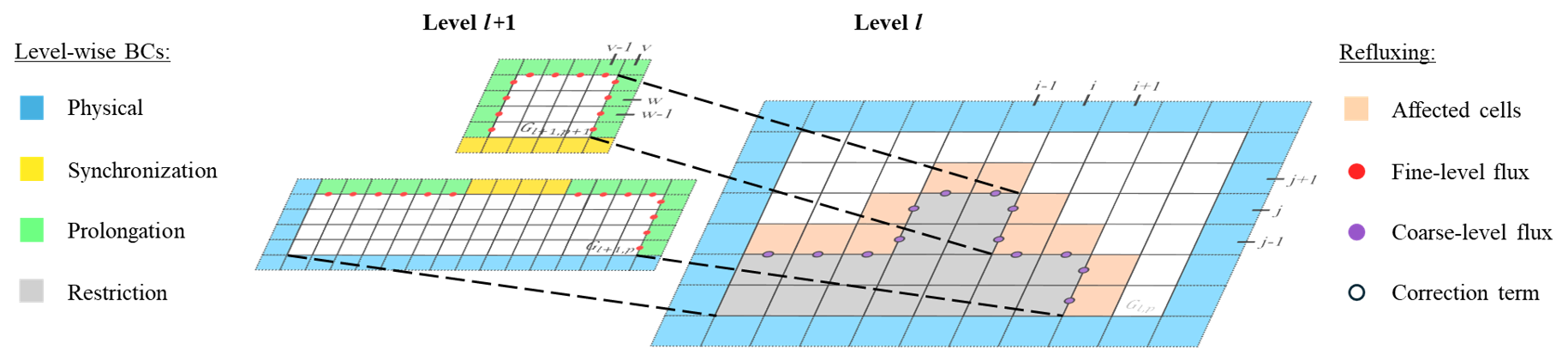}
\caption{Exemplary two-level grid layout with $r_l=2$ for the illustration of level-wise boundary conditions (BCs) and conservative refluxing. The cell coordinates of the coarser level are denoted with ($i$, $j$) and the finer level with ($v$, $w$), respectively \cite{strässle2025a-fully-conservative}.}
\label{fig:BCGrid}
\end{figure*}

A time-explicit finite-volume scheme in the form of a discrete velocity Boltzmann solver is used for the fully conservative discretization in space and time.
For simplicity, a first-order Euler-forward discretization in time was applied.
The resulting fully discretized system of hyperbolic \acp{PDE} is written as 
\begin{multline}
\label{eq:finalDiscretePDE}
    \{f_i,g_i\} (\bm{r}, t+ \delta t) = \{f_i,g_i\}(\bm{r},t) 
\\
    - \frac{\delta t}{\delta V} \sum_{\sigma \in \Theta}  \{\mathcal{F}_i,\mathcal{G}_i\} (\sigma, t)
    + \delta t \ \Omega_{\{f_i,g_i\}}(\bm{r},t)
    .  
\end{multline}
Here $\delta t$ is the time-step size, $\delta V$ the volume of the cell and $\{\mathcal{F}_i,\mathcal{G}_i\}$ fluxes through cell interfaces $\sigma$ of the cuboid surface $\Theta$. 
A nearest neighbor deformation (NND) interpolation scheme~\cite{NND1, NND2} was applied together with a generalized van Leer limiter~\cite{Harten1987Uniformly, Harten1987Uniformly}, which takes into account the ratio of successive slopes \cite{VanLeer_muscle, limiterRoe1986}, in order to compute fluxes \rebuttalUnmark{through} the interfaces in an accurate and stable manner by introducing more numerical dissipation if necessary.

More details on the kinetic model and discretization are elaborated in \cite{strässle2025consistent-compressible}.

\subsection{Adaptive mesh refinement methodology}

In this paper, the fully parallel, space-time adaptive conservative refinement methodology for finite-volume based discrete kinetic models is employed to showcase the refinement sensors.
Only the necessary concepts shall be highlighted hereafter.
For greater detail, the reader is referred to \cite{strässle2025a-fully-conservative}.

\subsubsection{Static refinement\label{sec:static-refinement}}

The methodology in the previous section described the evolution of the discrete kinetic system on a uniform Cartesian mesh.
Hereafter, a static \ac{2D} Cartesian domain is considered without loss of generality and the \ac{SAMR} approach is applied, as introduced and refined by Berger et al. \cite{Berger_Patchbased1, Berger_Patchbased2, Berger_Patchbased3, Berger_patchbased4} and intensively used in the \ac{CFD} and computational physics literature.
In this approach, the computational domain contains multiple levels $l \in \{0,L\}$, where the resolution is given by the level-wise refinement factor $r_{l} \geq 2$ with
\begin{equation}
\label{eq:refinementRatio}
    \frac{\delta x_{l}}{\delta x_{l+1}} = \frac{\delta y_{l}}{\delta y_{l+1}} = r_{l} 
    .
\end{equation}
The grid $G_{l,p}$ on each level is composed of patches $p \in \{0,P\}$.
Eq.~\eqref{eq:finalDiscretePDE} can be solved uniformly on every patch, due to the hyperbolicity of the system, with only minimal changes to the overall algorithm compared to a simple single level uniform grid.
These changes are summarized in the following and concern two distinct areas treating (A) information exchanges between patches and levels, and (B) the order of advancing levels in time.
For illustration, an exemplary two-level grid layout is depicted in Fig.~\ref{fig:BCGrid}.

(A) 
There are five types of information exchanges in total concerning a level $l$.
A halo of at least one ghost cell around each patch is required and needs to be filled with information, i.e. \acp{BC}, before advancing in time, which requires three types of \acp{BC} for ghost cells.
\begin{itemize}
\item[(1)] Physical \acp{BC}:  
Ghost cells overlapping a physical boundary need to be filled with the according physical boundary conditions. 
\item[(2)] Synchronization: 
For higher levels ($l>0$), ghost cells overlapping a cell of another patch of the same level (physical boundaries excluded) are populated with the corresponding values of the other patch.
\item[(3)] Prolongation:
For higher levels ($l>0$), ghost cells for which both above points do not apply are filled with interpolated values from underlying coarse cells on level $l-1$.
\end{itemize}
Additionally, another type of information exchange has to be supplied to level $l$, if a level $l+1$ exists.
\begin{itemize}
\item[(4)] Restriction:
Coarse cells on level $l$ which are covered by a patch of fine cells (excluding ghost cells) on level $l+1$ need to be filled by an average value of the fine cells.
This is done in order to populate the coarse cells with the most accurate values possible in case the fine-level patch is destroyed in the future, since the fundamental assumption in \ac{AMR} tells that a fine-level cell contains a more precise solution than the next coarser level cell. 
\end{itemize}
Furthermore, as there might exist a flux mismatch on the borders of a fine-level patch (excluding ghost cells) on level $l+1$ and the underlying coarse cell interfaces on level $l$, strict flux matching has to be enforced on level $l$ in order to retain the spirit of the strictly conservative finite-volume method across multiple levels.
\begin{itemize}
\item[(5)] \rebuttalUnmark{Conservative} refluxing: The flux mismatch is enforced using a correction pass, such that the original scheme, i.e. Eq.~\eqref{eq:finalDiscretePDE}, does not have to be modified locally but can still be applied uniformly.
Algorithmically this is done by using flux registers, which aggregate copies of the fluxes at the borders of a patch (excluding ghost cells) on level $l+1$ over space and time and overwrite the fluxes of underlying cells on level $l$ at these level-border locations.
\end{itemize}

\begin{figure}[t]
\centering
\includegraphics[width=0.99\linewidth]{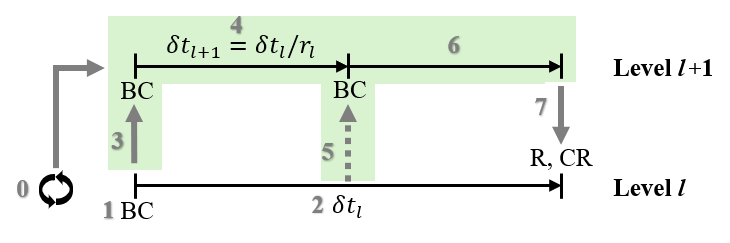}
\caption{Illustration of the updates and level-wise information exchanges in the subcycling approach.
BC refers to the boundary conditions of type physical, synchronization and prolongation, whereas R refers to restriction and CR to the \rebuttalUnmark{conservative} refluxing correction pass.
The horizontal axis marks time and the vertical axis displays the level $l$ and $l+1$ with an exemplary refinement ratio of $r_l = 2$. 
The order of operations is indicated with gray numbers, interpolations in space with gray arrows, 
and interpolation in space and time with gray dotted arrows, respectively. 
\rebuttalUnmark{The green shade marks operations conducted during recursion.}
In case a next finer level $l+2$ exists, the same recursive scheme would also be applied immediately before step number three and five \cite{strässle2025a-fully-conservative}.
}
\label{fig:Subcycling}
\end{figure}

(B)
A global time step over all levels is inefficient as the restriction on the \ac{CFL} number is determined by the smallest cell in the domain.
Another option is a recursive time integration routine, also referred to as subcycling.
By binding the level time steps to the refinement ratio \rebuttalUnmark{through} acoustic scaling, $\delta t \propto \delta x$, as 
\begin{equation}
    \frac{\delta t_{l}}{\delta t_{l+1}} = r_{l} 
    ,
\end{equation}
a constant CFL number across all levels can be ensured and greater numerical efficiency is obtained compared to diffusive scaling, $\delta t \propto \delta x^2$.
This particular subcycling routine, as applied in this work, is qualitatively depicted in Fig. \ref{fig:Subcycling}.
It illustrates the order of advancements in time and the application of information exchanges between levels.

More details and the explicit expressions for the information exchanges between levels are given in \cite{strässle2025a-fully-conservative}.
Further discussions and illustrations of \ac{SAMR} for generic \ac{FV} methods can be found in, e.g., \cite{Berger_Patchbased1, Berger_Patchbased2, Berger_Patchbased3, Berger_patchbased4, DeiterdingPhD, DeiterdingESAIM}.

\subsubsection{Adaptive refinement and parallel processing}

The adaptive component to a grid refinement methodology is given by a regridding routine.
It is applied every $n$-th level time step $\delta t_l$ and is responsible for creating, recreating or deleting patches, as well as modifying the underlying data structures.
A regridding routine can be summarized in a few steps.
\begin{itemize}
\item[(1)] Refinement sensors are applied to determine cells which require refinement. 
Note that this step, i.e. the development of suitable kinetic refinement sensors is the primary scope in this paper, and is therefore outlined separately in the next section.
\item[(2)] A clustering algorithm groups the tagged cells and forms rectangular patches.
A refinement efficiency, trading more neatly fitted patches for increased computational cost, and many other considerations such as problem specific buffer layers are taken into account.
Typical clustering algorithms employed in the CFD context are \cite{Berger_patchbased4, BergerRigoutsos}, however, more sophisticated ones can be found in the literature not limited to CFD and computational physics.
\item[(3)] The adapted grid layout is integrated into the level hierarchy and data structure.
\item[(4)] Cells in a newly created patch or in new region of a modified patch have to be filled with data, i.e. \acp{IC}. This is done by prolongation, analogous to the filling of \acp{BC} as outlined in Section~\ref{sec:static-refinement}.
\end{itemize}

Some aspects have to be considered for a parallel SAMR implementation on distributed memory machines.
On the one hand, a few changes have to be respected compared to the components discussed in Section~\ref{sec:static-refinement} concerning communication between patches and levels hosted on parallel machines.
On the other hand, parallel processing efficiency becomes a central challenge, as the grid layout consists of a complex and time-dependent hierarchy, where patches can move, appear and be destroyed, which can require lots of irregular memory referencing and redistribution of work load.
Different parallelization strategies exist, with considerations such as level-wise versus global domain decompositions, while minimal communication overhead and optimal load distribution is to be targeted.
Sophisticated load balancing algorithms control the distribution of patches between processors.
Usually a Knapsack algorithm or space-filling curves (Morton, Hilbert) are employed to connect cells as optimization strategy.
The parallel processing approach employed in this work, e.g., MPI+OMP/GPU, clustering, load balancing strategy and parallel scaling can be found in \cite{strässle2025a-fully-conservative}, with more background in the works of AMReX \cite{amrex1, amrex2, amrex3}.

\section{Refinement sensors\label{sec:Refinement-sensors}}

Refinement sensors are typically application-specific as they are primarily physically or heuristically motivated.
Physically motivated sensors usually consist of tracking macroscopic flow variables.
Other refinement criteria typically consist of an estimation of the order of the numerical scheme's truncation error by Richardson extrapolation, as, for example, applied in the original \ac{SAMR} implementations, cf. the series of works from Berger et al. \cite{Berger_Patchbased1, Berger_Patchbased2, Berger_Patchbased3, Berger_patchbased4, BergerRigoutsos}.

Macroscopic flow variables in the context of the present paper are all fields available from the continuum description of the fluid, i.e., conserved moments as well as some derived or primitive variables as, e.g., flow velocity, pressure, temperature, Mach number, vorticity, kinetic or internal energy.
These are typically employed with a norm of their value,
\begin{equation}
    \epsilon_{m} = \lvert m \rvert
    ,
\end{equation}
or the norm of their gradients,
\begin{equation}
    \epsilon_{\nabla m} = \lvert \nabla m \rvert
    ,
\end{equation}
where $m$ stands for the specific macroscopic variable of interest.
In this work, the Frobenius norm was applied.

In the following, 
\begin{itemize}
    \item[(1)] firstly, some local kinetic refinement sensors are introduced as a counterpart to the macroscopic sensors which are computed by gradients of the available macroscopic fields.
    \item[(2)] Secondly, some local kinetic refinement sensors are presented, which are not computable by the available macroscopic fields, hence are a unique feature of kinetic models.
\end{itemize}
A sensor for the quantity $n$ computed in the kinetic way shall be denoted as $\epsilon_{ {\rm kin. \ } n}$, whereas a sensor computed based on a macroscopic field shall be denoted as $\epsilon_{ {\rm mac. \ } n}$.
Note that, for the most part, kinetic sensors such as those presented below can be derived from the insights obtained in the \ac{CE} multiscale analysis, see \cite{strässle2025consistent-compressible}. 
The non-equilibrium distributions $\{f^{\rm neq},g^{\rm neq}\} = \{f,g\} - \{f^{\rm eq},g^{\rm eq}\}$ are hereafter approximated and referred to as the first-order expansions of the distributions in the \ac{CE} analysis, i.e $ \{f^{\rm neq},g^{\rm neq}\}  \approx \{f^{(1)},g^{(1)}\} $.
Also note that some of the sensors are model-specific, which is why a \ac{CE} analysis is indispensable.

\subsection{Kinetic sensors with macroscopic counterpart (class 1)}

In the following, some kinetic sensors are introduced which can replace a computation based on gradients of macroscopic variables.

\subsubsection{Viscous stresses sensor}
\label{sec:Sensor-Viscous-Stresses}

In addition to the conserved moments, other moment tensors can be monitored in kinetic models to gain meaningful physical insights.
The second-order tensor of $f$, i.e. the pressure tensor, containing all the flow contributions from pressure, shear and bulk dissipation is such an example.
The connection to the deviatoric Cauchy stress tensor of the NSF equations (i.e. the NS stress tensor) is given by 
\begin{equation}
    \bm{P}^{\rm neq}
    =       
    \sum_{i=0}^{Q-1} \bm{v}_i \otimes \bm{v}_i f_i^{\rm neq}
    = \bm{\tau}_{\rm NS} 
    .
\end{equation}
Therefore, the non-equilibrium pressure tensor can be employed as a sensor to measure the local strength of combined viscous momentum dissipation as
\begin{equation}
    \epsilon_{ {\rm kin.} \bm{\tau}_{\rm NS} }     
    =
    \left\lvert 
    \sum_{i=0}^{Q-1} \bm{v}_i \otimes \bm{v}_i f_i^{\rm neq}
    \right\rvert 
    .\label{eq:sensor-TauNS-kin}
\end{equation}
In comparison, the same sensor computed in a macroscopic way reads, cf. Eq. \eqref{eq:tauNS-1},
\begin{align}
    \epsilon_{ {\rm mac. } \bm{\tau}_{\rm NS} } 
    &= \left\lvert  
    - \mu (\bm{\nabla}\bm{u} + \bm{\nabla}\bm{u}^{\dagger}  )
    - \lambda (\bm{\nabla}\cdot\bm{u}) \bm{I} 
    \right\lvert 
\nonumber \\
    &= \left\lvert  
    - \mu \Bigl[ \bm{\nabla}\bm{u} + \bm{\nabla}\bm{u}^{\dagger} - \frac{2}{D} (\bm{\nabla}\cdot\bm{u}) \bm{I}\Bigr] 
    - \eta (\bm{\nabla}\cdot\bm{u}) \bm{I}
     \right\lvert  
    .\label{eq:sensor-TauNS-mac}
\end{align}

\subsubsection{Rate-of-compression sensor}
\label{sec:Sensor-Rate-of-compression}

Since the trace of $\bm{P}^{\rm neq}$ is evaluated as, 
\begin{equation}
    {\rm tr}(\bm{P}^{\rm neq}) 
    = 
    \sum_{i=0}^{Q-1} v_i^2 f_i^{\rm neq}
    =
    -
    \left(2\mu + D\lambda \right) (\bm{\nabla} \cdot \bm{u})
    ,
\end{equation}
a sensor accommodating the rate-of-compression tensor, can readily be computed,
\begin{align}
    \epsilon_{ {\rm kin.} \bm{C} } 
    &=                     
    \left\lvert  
    -
    \frac{1}{\left(2\mu + D\lambda \right) } 
    \left( \sum_{i=0}^{Q-1}  v_i^2  f_i^{\rm neq} \right) \bm{I}
    \right\rvert
    \nonumber \\
    &=                     
    \sqrt{D}
    \left\lvert  
    -
    \frac{1}{\left(2\mu + D\lambda \right) } 
    \left( \sum_{i=0}^{Q-1} v_i^2  f_i^{\rm neq} \right)
    \right\rvert
    ,\label{eq:sensor-C-kin}
\end{align}
where the Frobenius norm $|\bm{I}| = \sqrt{D}$.
In comparison, the same sensor computed in a macroscopic way reads, with cf. Eq. \ref{eq:rate-of-compression},
\begin{align}
    \epsilon_{ {\rm mac.} \bm{C} } 
    &=                     
    \left\lvert  
    \left( \bm{\nabla} \cdot \bm{u}\right) \bm{I}
    \right\rvert
    =      
    \sqrt{D}
    \left\lvert  
    \bm{\nabla} \cdot \bm{u}
    \right\rvert
   .\label{eq:sensor-C-mac}
\end{align}

\subsubsection{Rate-of-shear sensor}
\label{sec:Sensor-Rate-of-shear}

The rate-of-shear tensor can be expressed as
\begin{equation}
    \bm{S} = - \frac{1}{\mu} \bm{\tau_{\rm NS}} - \frac{\eta}{\mu} \bm{C}
    ,
\end{equation}
which results in a sensor written as 
\begin{multline}
    \epsilon_{ {\rm kin.} \bm{S} } 
    =
    \left\lvert  
    \frac{1}{\mu}
    \left[
    \frac{1}{D} 
    \left( 
    \sum_{i=0}^{Q-1}  v_i^2 f_i^{\rm neq}
    \right)
    \bm{I}
    -
    \left( 
    \sum_{i=0}^{Q-1} \bm{v}_i \otimes \bm{v}_i f_i^{\rm neq}
    \right)
    \right]
    \right\rvert 
    ,\label{eq:sensor-S-kin}
\end{multline} 
compared to, cf. Eq. \ref{eq:rate-of-shear},
\begin{equation}
    \epsilon_{ {\rm mac.} \bm{S} }
    = 
    \left\lvert  
    \bm{\nabla}\bm{u} + (\bm{\nabla}\bm{u})^{\dagger} 
    - \frac{2}{D} (\bm{\nabla}\cdot\bm{u}) \bm{I}
    \right\rvert  
    .\label{eq:sensor-S-mac}
\end{equation}

\subsubsection{Rate-of-strain sensor}
\label{sec:Sensor-Rate-of-strain}

Similarly, the rate of strain tensor can be found from Eq. \eqref{eq:tauNS-1} as 
\begin{equation}          
    \bm{E}
    =           
    \frac{1}{2}\bm{S} + \frac{1}{D}\bm{C}
    =           
    -\frac{1}{2\mu}\bm{\tau_{\rm NS}} - \frac{\lambda}{2\mu} \bm{C} 
    ,
\end{equation}
which results in a sensor written as 
\begin{multline}
    \epsilon_{ {\rm kin.} \bm{E} }  \rebuttalUnmark{=}
    \left\lvert  
    \frac{1}{2\mu}
    \left[ 
    \frac{\lambda}{D \eta } 
    \left( 
    \sum_{i=0}^{Q-1} v_i^2 f_i^{\rm neq}
    \right)
    \bm{I}
    -
    \left( 
    \sum_{i=0}^{Q-1} \bm{v}_i \otimes \bm{v}_i f_i^{\rm neq}
    \right)
    \right]
    \right\rvert 
    ,\label{eq:sensor-E-kin}
\end{multline} 
compared to, cf. Eq. \ref{eq:rate-of-strain},
\begin{equation}
    \epsilon_{ {\rm mac.} \bm{E} }
    = 
    \left\lvert  
    \frac{1}{2}\bm{\nabla}\bm{u} + \frac{1}{2}(\bm{\nabla}\bm{u})^{\dagger}
    \right\lvert  
    .\label{eq:sensor-E-mac}
\end{equation}

\subsubsection{Viscous heating sensor}
\label{sec:Sensor-Viscous-heating}

Another physically meaningful sensor based on moments of the distribution function can be applied to measure the viscous heating contribution to the energy flux.
The viscous heating vector, $q_{\rm H}= \bm{u}\cdot\bm{\tau}_{\rm NS}$, can directly be used as a sensor, computed as
\begin{multline}
    \epsilon_{ {\rm mac.} \bm{q_{\rm H}} }
    = 
    \left\lvert
    \bm{u}\cdot\bm{\tau}_{\rm NS}
     \right\rvert 
     = \left\lvert  
    - \mu (\bm{u}\cdot\bm{\nabla}\bm{u} + \bm{u}\cdot\bm{\nabla}\bm{u}^{\dagger} ) 
    - \lambda \bm{u}(\bm{\nabla}\cdot\bm{u})
    \right\lvert  
 \\ 
    = \left\lvert  
    - \mu \Bigl[ \bm{u}\cdot\bm{\nabla}\bm{u} + \bm{u}\cdot(\bm{\nabla}\bm{u})^{\dagger} - \frac{2}{D} \bm{u}(\bm{\nabla}\cdot\bm{u})\Bigr] 
    - \eta \bm{u}(\bm{\nabla}\cdot\bm{u})
     \right\lvert  
    .\label{eq:sensor-qH-mac}
\end{multline}
The kinetic sensor for the same quantity can be computed as
\begin{equation}
    \epsilon_{ {\rm kin.} \bm{q_{\rm H}} }
    = 
    \left\lvert
    \frac{\sum_{i=0}^{Q-1} \bm{v}_i f_i}{\sum_{i=0}^{Q-1} f_i}
    \cdot   
    \sum_{i=0}^{Q-1} \bm{v}_i \otimes \bm{v}_i f_i^{\rm neq} 
     \right\rvert 
    .\label{eq:sensor-qH-kin}
\end{equation}

\subsubsection{Energy flux sensor}
\label{sec:Sensor-Energy-flux}

A sensor for the energy flux $\bm{q}_{\rm NSF} =  \bm{q}_{\rm F} + \bm{q}_{\rm H}$, can be written as 
\begin{align}    \label{eq:sensor-qNSF-mac}
    \epsilon_{ {\rm mac.} \bm{q_{\rm NSF}} }
    &= 
    \left\lvert
    - \kappa \bm{\nabla}T
    + \bm{u}\cdot\bm{\tau}_{\rm NS}
    \right\rvert
    \\ \nonumber
    &= 
    \left\lvert
    - \kappa \bm{\nabla}T
    - \mu (\bm{u}\cdot\bm{\nabla}\bm{u} + \bm{u}\cdot(\bm{\nabla}\bm{u})^{\dagger} ) -  \lambda \bm{u}(\bm{\nabla}\cdot\bm{u})
    \right\rvert
    \\ \nonumber
    &{}\hspace{-1cm}
    = \left\lvert
    - \kappa \bm{\nabla}T
    - \mu \Bigl[ \bm{u}\cdot\bm{\nabla}\bm{u} + \bm{u}\cdot\bm{\nabla}\bm{u}^{\dagger} - \frac{2}{D} \bm{u}(\bm{\nabla}\cdot\bm{u})\Bigr] - \eta \bm{u}(\bm{\nabla}\cdot\bm{u})
    \right\rvert
    .\label{eq:sensor-qNSF-mac}
\end{align}
The formulation of a kinetic sensor is comparably simple, as the vector here termed $\bm{q}_{\rm NSF}$ is equal to $\bm{q}^{\rm neq}$.
The computation of $\bm{q}^{\rm neq}$, however, depends on the chosen energy split.
For the total energy split and the internal non-translational energy split, which are employed within this manuscript, the sensors can be found as 
\begin{equation}
        \rebuttalUnmark{ \epsilon_{ {\rm kin.} \bm{q_{\rm NSF}} } }
        =     
        \left\lvert
        \sum_{i=0}^{Q-1} \bm{v}_i g_i^{\rm neq} 
        \right\rvert
        ,\label{eq:sensor-qNSF-kin-tot}
\end{equation}
for the total energy split, and 
\begin{equation}
        \rebuttalUnmark{ \epsilon_{ {\rm kin.} \bm{q_{\rm NSF}} } }
            =
            \left\lvert
            \sum_{i=0}^{Q-1} \left( \bm{v}_i g_i^{\rm neq} + \bm{v}_i\frac{v_i^2}{2} f^{\rm neq} \right)
            \right\rvert
            ,\label{eq:sensor-qNSF-kin-ryk}
\end{equation}
for the internal non-translational split.

\subsubsection{Fourier heat flux sensor}
\label{sec:Sensor-Fourier-flux}

A macroscopic sensor for the Fourier heat flux can directly be written as 
\begin{equation}
    \epsilon_{ {\rm mac.} \bm{q_{\rm F}} }
    = 
    \left\lvert
    - \kappa \bm{\nabla}T
    \right\rvert
    .\label{eq:sensor-qF-mac}
\end{equation}
Mesoscopically, it can be found from 
$\bm{q}_{\rm F} = \bm{q}_{\rm NSF} - \bm{q}_{\rm H}$, 
where $\bm{q}_{\rm NSF} = \bm{q}^{\rm neq}$ as above. 
Therefore, an expression can be found as 
\begin{equation}
    \rebuttalUnmark{ \epsilon_{ {\rm kin.} \bm{q_{\rm F}} } }
    =     
    \left\lvert
    \sum_{i=0}^{Q-1} \bm{v}_i g_i^{\rm neq} 
    -
    \frac{\sum_{i=0}^{Q-1} \bm{v}_i f_i}{\sum_{i=0}^{Q-1} f_i}
    \cdot   
    \sum_{i=0}^{Q-1} \bm{v}_i \otimes \bm{v}_i f_i^{\rm neq} 
    \right\rvert
    ,\label{eq:sensor-qF-kin-tot}
\end{equation}
for the total energy split, and 
\begin{multline}
    \rebuttalUnmark{ \epsilon_{ {\rm kin.} \bm{q_{\rm F}} } }
    =
    \left\lvert
    \sum_{i=0}^{Q-1} \left( \bm{v}_i g_i^{\rm neq} + \bm{v}_i\frac{v_i^2}{2} f^{\rm neq} \right)
    \right. \\ \left. -
    \frac{\sum_{i=0}^{Q-1} \bm{v}_i f_i}{\sum_{i=0}^{Q-1} f_i}
    \cdot   
    \sum_{i=0}^{Q-1} \bm{v}_i \otimes \bm{v}_i f_i^{\rm neq} 
    \right\rvert
    ,\label{eq:sensor-qF-kin-ryk}
\end{multline}
for the internal non-translational split, respectively.

\subsection{Purely kinetic sensors (class 2)}

Hereafter, some purely kinetic sensors are introduced, which cannot be recreated based on macroscopic variables.

\subsubsection{Knudsen sensor}
\label{sec:Sensor-Knudsen}

A local estimate of the lattice Knudsen number, as proposed by \citet{Thorimbert_KNsensor}, is considered in this manuscript for comparison purposes.
In the \ac{CE} multiscale analysis, the Knudsen number is related to the smallness parameter.
For an expansion up to order one with respect to the smallness parameter/Knudsen number,
$f = f^{(0)} + {\rm Kn} f^{(1)} + \mathcal{O}({\rm Kn}^2)$,
and again referring to the non-equilibrium distribution as
$f^{\rm neq} = f-f^{\rm eq} \approx f^{(1)}$,
the relation
\begin{equation}
    {\rm Kn} \propto \frac{f^{(1)}}{f^{(0)}} \approx \frac{f^{\rm neq} }{f^{\rm eq}}
\end{equation}
holds.
Hence, an estimate of the order of the Knudsen number can be gained by an average over all $Q$ discrete velocities, with the sensor written as
\begin{equation}
    \epsilon_{{\rm kin.Kn\text{\footnotesize{\cite{Thorimbert_KNsensor}}}}}
    = \frac{1}{Q} \sum_{i=0}^{Q-1} 
    \left\lvert\frac{f^{\rm neq}_i}{f_i^{\rm eq}}\right\rvert 
    = \frac{1}{Q} \sum_{i=0}^{Q-1} \left\lvert \frac{f_i - f_i^{\rm eq}}{f_i^{\rm eq}}
    \right\rvert
    .\label{eq:sensor-Knudsen-kin-THORIMBERT}
\end{equation}
This sensor has been used for example in \cite{Coreixas2020adaptivevelocity} for \ac{AAR} by introducing a dynamic relaxation time into the BGK collision operator (locally added dissipation to damp phenomena related to departures from equilibrium) when the sensor value is too high, i.e. when the recovery of the proper macroscopic behavior is at risk due to hydrodynamic limit considerations.

Within this manuscript, the non-averaged expression is used. 
Furthermore, as a double-distribution framework is employed, both contributions from the $f$- and $g$-distribution can be taken into account.
For this, a sensor is constructed in the simplest form by superposition of the individual sensors for the $f$- and $g$-distribution.
The sensor then reads
\begin{equation}
    \epsilon_{{\rm kin.Kn}} = \epsilon_{{\rm kin.Kn(f)}} + \epsilon_{{\rm kin.Kn(g)}}
    ,\label{eq:sensor-Knudsen-kin}
\end{equation}
where
\begin{gather}
    \epsilon_{{\rm kin.Kn(f)}}
    = \sum_{i=0}^{Q-1} 
    \left\lvert \frac{f^{\rm neq}_i}{f_i^{\rm eq}} \right\rvert 
    = \sum_{i=0}^{Q-1} 
    \left\lvert \frac{f_i - f^{\rm eq}_i}{f_i^{\rm eq}} \right\rvert 
    ,\label{eq:sensor-Knudsen-kin_f}
\\
    \epsilon_{{\rm kin.Kn(g)}}
    = \sum_{i=0}^{Q-1} 
    \left\lvert \frac{g^{\rm neq}_i}{g_i^{\rm eq}} \right\rvert 
    = \sum_{i=0}^{Q-1} 
    \left\lvert \frac{g_i - g^{\rm eq}_i}{g_i^{\rm eq}} \right\rvert 
 .\label{eq:sensor-Knudsen-kin_g}
\end{gather}
Note that this form of constructing a sensor by superposition in order to account for both distributions will also be employed for the remainder of this manuscript.

\subsubsection{Non-equilibrium sensor}
\label{sec:Sensor-Neq}

Measuring the strength of non-equilibrium effects in the flow, i.e. the departure from equilibrium, can be done by the pure non-equilibrium contributions.
Note that the non-equilibrium contribution $f_i^{\rm neq}$ to the distribution function is expected to grow in the presence of local density, velocity and temperature gradients, as apparent by the \ac{CE} expansion and the invariance defect of the \ac{MB} distribution \cite{hosseini2023lattice}.
A non-equilibrium sensor, involving contributions from both distribution functions, can directly be written as 
\begin{multline}
    \epsilon_{{\rm kin.neq}}
     =  \sum_{i=0}^{Q-1}  \left\lvert f_i^{\rm neq}  \right\rvert + \sum_{i=0}^{Q-1} \left\lvert g_i^{\rm neq}  \right\rvert
 \\
     = \sum_{i=0}^{Q-1}  \left\lvert f_i - f_i^{\rm eq}  \right\rvert 
     + \sum_{i=0}^{Q-1}\left\lvert g_i - g_i^{\rm eq}  \right\rvert 
    .\label{eq:sensor-neq-kin}
\end{multline}
This expression evaluates to zero at equilibrium, clearly highlighting regions with non-equilibrium contributions.

\subsubsection{Quasi- and shifted equilibrium sensors}
\label{sec:Sensor-QE-Neq-Kn}

The same idea can be applied for models containing quasi-equilibrium populations, such as the one treated in this manuscript for non-unity Prandtl number.
By recalling the picture of the total relaxation process $f \rightarrow f^* \rightarrow f^{\rm eq}$, one can compare individual relaxation processes with each other and measure the strength of local \ac{QE} contributions.
For this, both relaxation steps $f \rightarrow f^*$ and  $f^* \rightarrow f^{\rm eq}$ can be evaluated.
Herein, the relaxation process $f^* \rightarrow f^{\rm eq}$ is targeted, as it vanishes in the limit of the single-relaxation BGK kinetic model. 
I.e., the purpose is to measure the local deviation from the single-relaxation BGK kinetic model by looking at the local dominance of fluxes introduced \rebuttalUnmark{through} the intermediate quasi-equilibrium state.
Again, a simple superposition of the individual values stemming from the $f$- and $g$-distribution is applied.
A sensor quantifying the off-equilibrium part with respect to the \ac{QE} populations can be written in the form of Eq. \ref{eq:sensor-neq-kin} as
\begin{equation}
    \epsilon_{{\rm kin.qe}}
    = \sum_{i=0}^{Q-1}\left\lvert f_i^* -  f_i^{\rm eq} \right\rvert  
    + \sum_{i=0}^{Q-1} \left\lvert g_i^* -  g_i^{\rm eq} \right\rvert  
    .\label{eq:Sensor-qe-kin}
\end{equation}
A related sensor could also be expressed in the style of the lattice $\mathrm{Kn}$ number as in Eqs. \ref{eq:sensor-Knudsen-kin} to \ref{eq:sensor-Knudsen-kin_g}, by normalizing the expression with the equilibrium state.

It shall be noted that such a quasi-equilibrium sensor can be applied to all sorts of quasi-equilibria and shifted-equilibria attractors, in order to measure the local contributions from these populations.
Such examples are, for instance, multicomponent-models using \ac{QE} for non-unity Schmidt numbers,
or models for non-ideal fluids using shifted-equilibria for the application of local forcing terms as, e.g., in \cite{Hosseini2022Towards, Karlin2025Practical, hosseini_2025_compressiblenonideal}.

\subsubsection{Relative-$H$ sensors}
\label{sec:Relative-H-sensor}

For some flow situations, knowledge about a local entropy measure might be of interest.

\rebuttalUnmark{Through} the relation of entropy and the equilibrium $H$-function in Eq. \ref{eq:relation-H-entropy}, it is apparent that a sensor for the equilibrium entropy could be constructed in both the kinetic and macroscopic way; 
explicitly, using the discrete $H$-function of Eq. \eqref{eq:H-function-discrete} at equilibrium, 
\begin{equation}
    H(f^{\rm eq}) =  \sum_{i=0}^{Q-1} f_i^{\rm eq} \ln{\frac{f_i^{\rm eq}}{w_i}} 
    ,\label{eq:eq-entropy-kin}
\end{equation}
marking negative scaled equilibrium entropy,
and using the Gibbs relations
\begin{equation}
    s 
    = C_v \ln{\frac{T}{\rho^{\gamma-1}}} 
    \underbrace{
    + s_0 - C_v \ln{\frac{T_0}{\rho_0^{\gamma-1}}}
    }_{\mathcal{C}_0} 
    = C_v \ln{\frac{p}{R \rho^{\gamma}}}
    \underbrace{
    + s_0 - C_v \ln{\frac{p_0}{R \rho_0^{\gamma}}}
    }_{\mathcal{C}_0} 
    ,\label{eq:gibbs-entropy-macro}
\end{equation}
where the constant $\mathcal{C}_0$ containing all contributions from a reference state $\rho_0$, $p_0$, $T_0$, $s_0$ can conveniently be evaluated at absolute zero, marking an expression for equilibrium entropy.
Another measure, which cannot be reproduced macroscopically, would be a sensor for the full (negative, scaled) off-equilibrium entropy, which can be constructed using the discrete $H$-function evaluated at the local distribution, as in Eq. \eqref{eq:H-function-discrete},
\begin{equation}
        H(f) = \sum_{i=0}^{Q-1} f_i \ln{\frac{f_i}{w_i}}
        .\label{eq:off-entropy-kin}
\end{equation}
However, the above expressions would not be particularly useful for the use as refinement sensors, as they depict absolute values of an entropy measure.

More interesting would be an expression for (negative, scaled) relative entropy, i.e. entropy measured relative to an equilibrium value.
This can be expressed for example as in \cite{Brownlee2006Stabilization}, by evaluating the difference of the discrete $H$-function evaluated at the local distribution and the local equilibrium, constructing a sensor as
\begin{multline}
    \epsilon_{{\rm kin}.H^{\rm R}}
    = \left\lvert H(f) -  H(f^{\rm eq}) \right\rvert + \left\lvert  H(g) -  H(g^{\rm eq}) \right\rvert
\\
    =  \left\lvert 
        \sum_{i=0}^{Q-1} f_i\ln{\frac{f_i}{w_i}} 
        - \sum_{i=0}^{Q-1} f_i^{\rm eq} \ln{\frac{f_i^{\rm eq}}{w_i}} 
    \right\rvert
\\
    +  \left\lvert 
        \sum_{i=0}^{Q-1} g_i\ln{\frac{g_i}{w_i}} - \sum_{i=0}^{Q-1} g_i^{\rm eq} \ln{\frac{g_i^{\rm eq}}{w_i}} 
    \right\rvert
    ,\label{eq:Hneq-sensor}
\end{multline}
again involving contributions from both distribution functions.
Another option for a (negative, scaled) relative entropy can be formulated in the form of a \ac{KL} divergence between the distribution and its equilibrium \cite{Dzanic_2025_TG_Boltzmann, Brownlee2008Nonequilibrium}, resulting in the expression
\begin{equation}
     H^{\rm KL}(f, f^{\rm eq})  
    =  \sum_{i=0}^{Q-1} f_i \ln{\frac{f_i}{f_i^{\rm eq}}}
    .\label{eq:relative-entropy-kin}
\end{equation}
The reader is referred to, e.g., \cite{Dzanic_2025_TG_Boltzmann} for a derivation on the basis of a discrete (negative, scaled) cross-entropy.
A sensor can then be constructed, again superposing expressions for both distribution functions, as
\begin{equation}
        \epsilon_{ {\rm kin.} H^{\rm KL} }    
    = \left\lvert \sum_{i=0}^{Q-1} f_i \ln{\frac{f_i}{f_i^{\rm eq}}}\right\rvert
    + \left\lvert \sum_{i=0}^{Q-1} g_i \ln{\frac{g_i}{g_i^{\rm eq}}}\right\rvert
    .\label{eq:HKL-sensor}
\end{equation}
This can be understood as a measure of distance between the distribution and its equilibrium.
Note that these relative $H$-sensors, i.e. $H^{\rm R}$ and $H^{\rm KL}$ in \ref{eq:Hneq-sensor} and \ref{eq:HKL-sensor}, evaluate to zero at equilibrium and therefore mark useful sensors compared to the expressions in Eqs. \ref{eq:eq-entropy-kin} to \ref{eq:off-entropy-kin}, which do not measure relative entropy and thus have a non-zero baseline.

\subsubsection{$H$-dissipation sensors}
\label{sec:Relative-H-dissipation-sensor}

From the $H$-theorem, cf. Eqs. \eqref{eq:Htheorem} and \eqref{eq:sigma-function-discrete}, it becomes apparent that a look into the production rate of entropy, or dissipation rate of $H$, might also be beneficial, in order to quantify the local strength of entropy production / $H$-dissipation.
The discrete dissipation of the $H$-function in Eq. \eqref{eq:sigma-function-discrete} can readily be used to construct a sensor as $-\sigma_H  = |\sigma_H| \geq 0$. 
The sensor reads
\begin{equation}
    \epsilon_{ {\rm kin.} \sigma_H }
    = - \left(
    \sum_{i=0}^{Q-1} \Omega_{f_i} \ln{\frac{f_i}{w_i}}
    + \sum_{i=0}^{Q-1} \Omega_{g_i} \ln{\frac{g_i}{w_i}}
    \right)
    ,\label{eq:Dissip-H-sensor}
\end{equation}
accounting for both distributions.
Similarly, a sensor for the dissipation rate of the (negative, scaled) relative entropy introduced before can be constructed, evaluating to
\begin{equation}
    \sigma_{H^{\rm KL}} (f, f^{\rm eq}) = \sum_{i=0}^{Q-1} \Omega_{f_i} \ln{\frac{f_i}{f_i^{\rm eq}}}
    ,\label{eq:discrete-sigma-HKL}
\end{equation}
which yields
\begin{equation}
    \epsilon_{ {\rm kin.} \sigma_{H^{\rm KL}} } 
    = - \left(\sum_{i=0}^{Q-1} \Omega_{f_i} \ln{\frac{f_i}{f_i^{\rm eq}}}
    +   \sum_{i=0}^{Q-1} \Omega_{g_i} \ln{\frac{g_i}{g_i^{\rm eq}}}
    \right)
    .\label{eq:Dissip-HKL-sensor}
\end{equation}
For both sensors, $\{ \Omega_{f_i}, \Omega_{g_i} \}$ denotes the discrete collision operator for the $f_i$ and $g_i$ populations.
Note that any BGK type collision can be applied here, however, for the content of this manuscript, the discrete QE-BGK is used as in Eq. \eqref{new_collision_fi}.

\subsubsection{Entropic estimate sensor}
\label{sec:Sensor-AlphaPoly}

The notion of the entropic estimate $\alpha$, as used in the non-linear stabilization scheme of the \ac{ELBM} \cite{Karlin_entropicEQ, Karlin_Gibbs, Ali_reviewEntropic}, can also be introduced as a sensor for \ac{AMAR}.
The entropic estimate $\alpha$ measures the value which satisfies the condition of equal entropy of the mirror state by means of the discrete $H$-function, cf. Eq. \eqref{eq:H-function-discrete},
i.e. it is found as the positive root of
\begin{equation}
        H(f + \alpha_f(f^{\rm eq} - f)) 
        =H(f)
        .\label{eq:H-mirr-equal}
\end{equation}
In essence, the entropic estimate measures the maximum available over-relaxation path in an LB scheme without violation of the $H$-theorem. 
The rationale behind this sensor is to measure when the recovery of the proper macroscopic behavior is at risk due to potential violations of the $H$-theorem.
Recall that the standard \ac{LBGK} mirror state has a fixed value of $\alpha_f = 2$.
Hence, with this idea in mind, one can construct the sensor 
\begin{equation}
    \epsilon_{{\rm kin.}\alpha} 
    = \left\lvert 2 - \alpha_f \right\rvert 
    + \left\lvert 2 - \alpha_g \right\rvert
    ,\label{eq:sensor-alpha-kin}
\end{equation}
including again the contributions from both distribution functions.
Note that, while the contributions from $f$ and $g$ are superposed herein for the purpose of constructing a sensor, usage of the entropic estimate in an ELBM within a double-distribution framework can be done as, e.g., in \cite{2016Entropic}.

A preliminary assessment showed that the polynomial approximation of the entropic estimate is sufficiently accurate for the application to \ac{AMAR} and can save significant amounts of computational effort compared to, e.g., the options of a Newton--Raphson iterative solver or bisection search algorithm.
The polynomial approximation is written as \cite{Ali_reviewEntropic}
\begin{equation}
    \alpha_{\{f,g\}} = 2 - 4 \frac{a_2}{a_1} + 16 \frac{a_2^2}{a_1^2} - 8 \frac{a_3}{a_1} + 80 \frac{a_2 a_3}{a_1^2} - 80 \frac{a_2^3}{a_1^3} - 16 \frac{a_4}{a_1}
    ,\label{eq:alpha-polynomial-1}
\end{equation}
with 
\begin{equation}
    a_{n\{f,g\}} = \frac{(-1)^{n-1}}{n(n+1)} \sum_{i=0}^{Q-1} \left\{ \frac{{(f_i^{\rm neq})}^{n+1}}{f_i^n} , \frac{{(g_i^{\rm neq})}^{n+1}}{g_i^n} \right\}
    .\label{eq:alpha-polynomial-2}
\end{equation}

\subsection{Overview and further considerations}

\begin{table*}[t]
\footnotesize
\centering
\newcolumntype{P}[1]{>{\centering\arraybackslash}p{#1}}
\newcolumntype{L}[1]{>{\raggedright\arraybackslash}p{#1}}
\begin{tabular}{|l|l|l|l|l|l|}
\hline
Class & Name & Type & Variable & Section & Equation(s)  \\
\hline
\hline
\multirow{13}{*}{\rotatebox{90}{\parbox{4.5cm}{\centering Class 1:\\Kinetic with macroscopic counterpart}}}
    &\multirow{2}{*}{Viscous stresses sensor} &Macroscopic &$\epsilon_{ {\rm mac.} \bm{\tau}_{\rm NS} }$    &\multirow{2}{*}{\ref{sec:Sensor-Viscous-Stresses}}&\eqref{eq:sensor-TauNS-mac}\\
    &&Kinetic &$\epsilon_{ {\rm kin.} \bm{\tau}_{\rm NS} }$ & &\eqref{eq:sensor-TauNS-kin}\\
    \cline{2-6}
    &\multirow{2}{*}{Rate-of-compression sensor} &Macroscopic &$\epsilon_{ {\rm mac.}\bm{C}}$ &\multirow{2}{*}{\ref{sec:Sensor-Rate-of-compression}}&\eqref{eq:sensor-C-mac}\\
    &&Kinetic &$\epsilon_{ {\rm kin.} \bm{C} } $ & &\eqref{eq:sensor-C-kin}\\
    \cline{2-6}
    &\multirow{2}{*}{Rate-of-shear sensor} &Macroscopic &$\epsilon_{ {\rm mac.}\bm{S}}$&\multirow{2} {*}{\ref{sec:Sensor-Rate-of-shear}}&\eqref{eq:sensor-S-mac}\\
    &&Kinetic &$\epsilon_{ {\rm kin.} \bm{S} } $ & &\eqref{eq:sensor-S-kin}\\
    \cline{2-6}
    &\multirow{2}{*}{Rate-of-strain sensor} &Macroscopic &$\epsilon_{ {\rm mac.}\bm{E}}$&\multirow{2}{*}{\ref{sec:Sensor-Rate-of-strain}}&\eqref{eq:sensor-E-mac}\\
    &&Kinetic &$\epsilon_{ {\rm kin.} \bm{E} } $ & &\eqref{eq:sensor-E-kin}\\
    \cline{2-6}
    &\multirow{2}{*}{Viscous heating sensor} &Macroscopic &$\epsilon_{ {\rm mac.} \bm{q_{\rm H}}}$ &\multirow{2}{*}{\ref{sec:Sensor-Viscous-heating}} & \eqref{eq:sensor-qH-mac}\\
    &&Kinetic &$\epsilon_{ {\rm kin.} \bm{q_{\rm H}} }$ & &\eqref{eq:sensor-qH-kin}\\
    \cline{2-6}
    &\multirow{2}{*}{Energy flux sensor} &Macroscopic &$\epsilon_{ {\rm mac.} \bm{q_{\rm NSF}}}$ &\multirow{2}{*}{\ref{sec:Sensor-Energy-flux}}&\eqref{eq:sensor-qNSF-mac}\\
    &&Kinetic &$\epsilon_{ {\rm kin.} \bm{q_{\rm NSF}} }$ & &\eqref{eq:sensor-qNSF-kin-tot} / \eqref{eq:sensor-qNSF-kin-ryk} \rebuttalUnmark{$^\star$} \\
    \cline{2-6}
    &\multirow{2}{*}{Fourier heat flux sensor} &Macroscopic &$\epsilon_{ {\rm mac.} \bm{q_{\rm F}} }$ &\multirow{2}{*}{\ref{sec:Sensor-Fourier-flux}}&\eqref{eq:sensor-qF-mac}\\
    &&Kinetic &$\epsilon_{ {\rm kin.} \bm{q_{\rm F}} }$ & &\eqref{eq:sensor-qF-kin-tot} / \eqref{eq:sensor-qF-kin-ryk} \rebuttalUnmark{$^\star$} \\ 
\hline
\hline
\multirow{9}{*}{\rotatebox{90}{\parbox{3cm}{\centering Class 2:\\Purely kinetic}}}
    &Knudsen sensor &Kinetic &$\epsilon_{{\rm kin.Kn}}$ &\ref{sec:Sensor-Knudsen}&\eqref{eq:sensor-Knudsen-kin}\\
    \cline{2-6}
    &Non-equilibrium sensor &Kinetic &$\epsilon_{{\rm kin.neq}}$ &\ref{sec:Sensor-Neq}&\eqref{eq:sensor-neq-kin}\\
    \cline{2-6}
    &Quasi- and shifted-equilibrium sensors &Kinetic &$\epsilon_{{\rm kin.qe}}$&\ref{sec:Sensor-QE-Neq-Kn}&\eqref{eq:Sensor-qe-kin}\\
    \cline{2-6}
    &\multirow{2}{*}{Relative-$H$ sensors} &Kinetic &$\epsilon_{ {\rm kin.} H^{\rm R} }$ &\multirow{2}{*}{\ref{sec:Relative-H-sensor}}&\eqref{eq:Hneq-sensor}\\
    &&Kinetic &$\epsilon_{ {\rm kin.} H^{\rm KL} }$ & &\eqref{eq:HKL-sensor}\\
    \cline{2-6}
    &\multirow{2}{*}{Relative-$H$-dissipation sensors} &Kinetic &$\epsilon_{ {\rm kin.} \sigma_{H}}$ &\multirow{2}{*}{\ref{sec:Relative-H-dissipation-sensor}}&\eqref{eq:Dissip-H-sensor}\\
    &&Kinetic&$\epsilon_{ {\rm kin.} \sigma_{H^{\rm KL}} }$ & &\eqref{eq:Dissip-HKL-sensor}\\
    \cline{2-6}
    &Entropic estimate sensor &Kinetic &$\epsilon_{{\rm kin.}\alpha}$ &\ref{sec:Sensor-AlphaPoly}&\eqref{eq:sensor-alpha-kin}, \eqref{eq:alpha-polynomial-1}, and \eqref{eq:alpha-polynomial-2}\\
    \hline
\end{tabular}
\caption{
Summary of refinement sensors from Section~\ref{sec:Refinement-sensors} applied in this manuscript.
\rebuttalUnmark{Superscript "$^\star$" indicates dependence on the energy split.}
}
\label{table1}
\end{table*}

A few considerations are in order here.

From a mathematical perspective, positivity of the equilibria populations is a requirement to evaluate the entropy-related ($H$-function) sensors introduced in previous sections,
which is, however, not strictly given by the use of Grad--Hermite expansions, as employed within this manuscript.
No issues related to non-positivity of the populations were discovered for the evaluations and applications discussed in the remainder of this manuscript.
For the purpose of demonstration of these quantities within the use as refinement sensors, the Grad--Hermite expansion was considered sufficient herein. 
It shall be noted, however, that a change to the entropic equilibria construction approach (together with a numerical evaluation of $\alpha$, cf. Eq. \eqref{eq:alpha-polynomial-1}) might be needed in order to strictly fulfill this requirement in case these sensors fail. 

Moreover, the list of refinement sensors introduced in the previous sections is by no means exclusive. Some comments are in order here.

Most of the local kinetic sensors, which are herein applied to account for both distributions, can readily be used for a single distribution function.
Furthermore, many different formulations of the same idea can be applied as a sensor, for example, evaluating the difference between the distribution and its equilibrium for $H$-sensors in a slightly varying way, or accounting for the effects by both distribution functions in a different way then by superposition. 

Furthermore, much effort has been spent in the classical \ac{CFD} community to find more sophisticated macroscopic sensors for specific types of flows.
Accordingly, kinetic analogues can be identified for these sensors, with some of them requiring modifications to the kinetic model.
An example is the dilatation dissipation rate, defined as, $D_{\rm dil} = \mu (\bm{\nabla}\cdot \bm{u})^2$, which can also be readily be computed the kinetic way through the rate-of-compression tensor $\bm{C}$.
Other macroscopic quantities such as the vorticity vector, defined as  $\bm{\omega} = \bm{\nabla} \times \bm{u}$, could be applied as a sensor.
This, however, requires adaptation of the kinetic model:
Approaches for the computation of vorticity based on distribution functions have been introduced by, e.g., \cite{localKineticVorticityMRT} for MRT models and \cite{localKineticVorticityDDF} for \ac{DDF} models. 
These approaches generally consist of introducing anisotropy in higher-order equilibrium moments which are not relevant to the macroscopic hydrodynamics in order to obtain the anti-symmetric components of $\bm{\nabla \bm{u}}$.
Another quantity, the solenoidal component of viscous dissipation, $\sigma_{\rm sol} = \mu \bm{\omega} \cdot \bm{\omega}$, can also be obtained this way.
Furthermore, by decomposition of the velocity gradient tensor 
$\bm{\nabla}\bm{u}$ 
into a symmetric and asymmetric part as 
$\nabla\bm{u} = \bm{E} + \bm{\Omega}$, 
where the asymmetric rate-of-rotation tensor, or vorticity tensor, is given by 
$ \bm{\Omega} = \bm{\nabla}\bm{u} - \bm{\nabla}\bm{u}^\dagger $, 
the $Q$-criterion can be obtained as
$\epsilon_{{\rm mac.Qcrit}} = \frac{1}{2} \left( \left\lvert  \bm{\Omega} \right\rvert ^2 - \left\lvert \bm{E} \right\rvert ^2 \right)$.
The $Q$-criterion displays vorticity dominated areas of the flow field by positive values, while negative values indicate areas dominated by the strain rate. 
Note the relation of the rate-of-rotation tensor and the vorticity vector via the Frobenius norm as $\left\lvert  \bm{\Omega} \right\rvert = \sqrt{2} \left\lvert \bm{\omega}\right\rvert$.
Hence, an expression for the kinetic analogues of both the solenoidal viscous dissipation and the $Q$-criterion, even in other formulations such as the objective $Q$-criterion in \cite{pedergnana2025objectiveqcriterion, Kogelbauer_Pedergnana_2026}, can be obtained this way via the route of the vorticity vector.

However, with these considerations and examples being mentioned, the validation and applications in the remainder of this paper are restricted to what has been introduced in the previous sections. 
A summary of all sensors validated and applied in this paper is given in Table~\ref{table1}.

\section{Evaluation of the local kinetic sensors\label{sec:Validation}}

All kinetic sensors summarized in Table~\ref{table1} are evaluated without the application of AMAR.
They are compared to their macroscopic counterpart in case of class 1 sensors.
The application of AMR based on these sensors follows in the Section~\ref{sec:ApllicationAMR}.
In all simulations, unless otherwise stated, the parameters for the presented results were set to those of Air, i.e. \mbox{$\gamma=1.4$} and \mbox{$\mathrm{Pr}=0.71$} with the gas constant set to $R=1$.
\Acp{IC} were supplied by means of equilibrium populations.
\Ac{1D} tests were run as pseudo-1D with periodic \acp{BC} in the pseudo direction, otherwise von Neumann \acp{BC} were applied.
The time step was estimated with
\begin{equation}
    \delta t = \frac{\min(\delta x,\delta y)}{|{\bm{u}| + c_s}} \text{CFL}
    ,
\end{equation}
where the CFL number was set to $0.01$ in order to also stably simulate higher Ma numbers.
Central finite differences were used for the computation of the macroscopic sensors involving gradients (class 1).
Physical quantities are reported as non-dimensional values whenever no units are mentioned.
Note again that a kinetic model recovering the \ac{NSF} equations in the hydrodynamic limit was used for all test cases run herein, despite some of the cases being known from the literature as Euler level test cases; 
these test cases were adopted since the identification of locations with, e.g., high viscous and thermal stresses or high entropy production is rather intuitive based on the solutions of macroscopic variables.
As the solver was already validated in \cite{strässle2025consistent-compressible, strässle2025a-fully-conservative} on all test cases discussed in this manuscript,  no further validation of the solver and comparison with reference solution is discussed herein for the sake of readability.
Lastly, note that the rate-of-compression, rate-of-shear and rate-of-strain are proportional to the Navier--Stokes stress tensor in \ac{1D} and are therefore not shown and discussed for the (pseudo) one-dimensional simulations.

\subsection{Sod shock tube\label{sec:val-Sod-tube}}

First, a \ac{1D} validation is conducted using the Sod shock tube \cite{Sodtube} as the classical benchmark for compressible flows.
The case was initialized as
\begin{equation}
  (\rho, p, u_x) =
    \begin{cases}
      (1, 1, 0), & 0 \leq x \leq 0.5,\\
      (0.125, 0.1, 0), & 0.5 < x \leq 1,
    \end{cases}       
\end{equation}
in a domain $x \in [0,1]$.
A dynamic shear viscosity of \mbox{$\mu = \num{5e-5}$} was imposed and the simulation was evaluated at a time $t_{\rm end} = 0.2$s.

\begin{figure*}
    \centering
    \includegraphics[height=0.95\textheight, keepaspectratio]{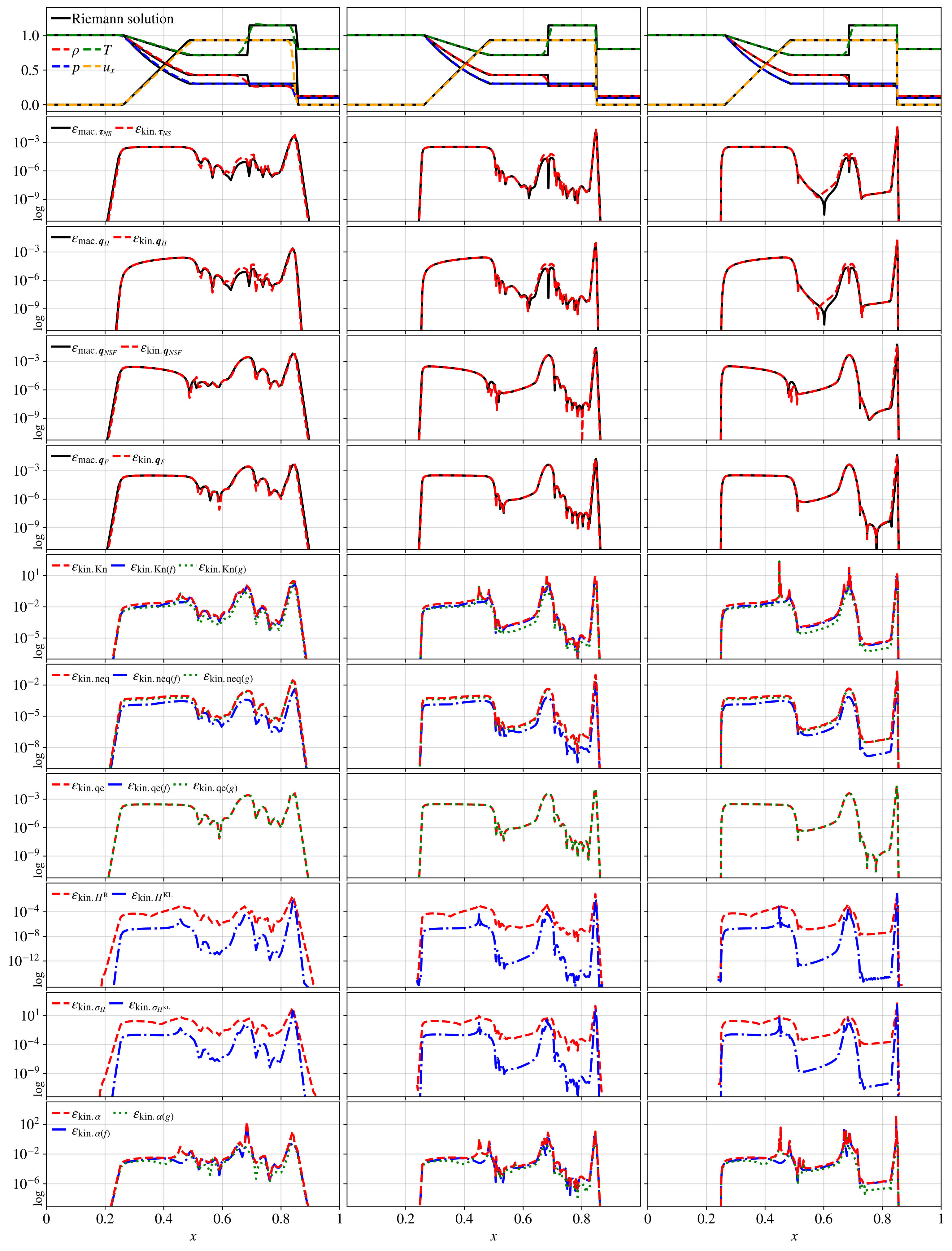}%
    \caption{
    Sensors for the Sod shock tube using the 
    total energy split
    for different resolutions depicted from left to right; $\delta x = L_x/128$, $\delta x = L_x/512$ and $\delta x = L_x/2048$.
    The legends hold for the whole row. 
    Note that the $y$-axis of the sensors are logarithmic, which is indicated on the respective axis for convenience. 
    }
    \label{fig:Val_Sod_Tot}%
\end{figure*}

\begin{figure*}
    \centering
    \includegraphics[height=0.95\textheight, keepaspectratio]{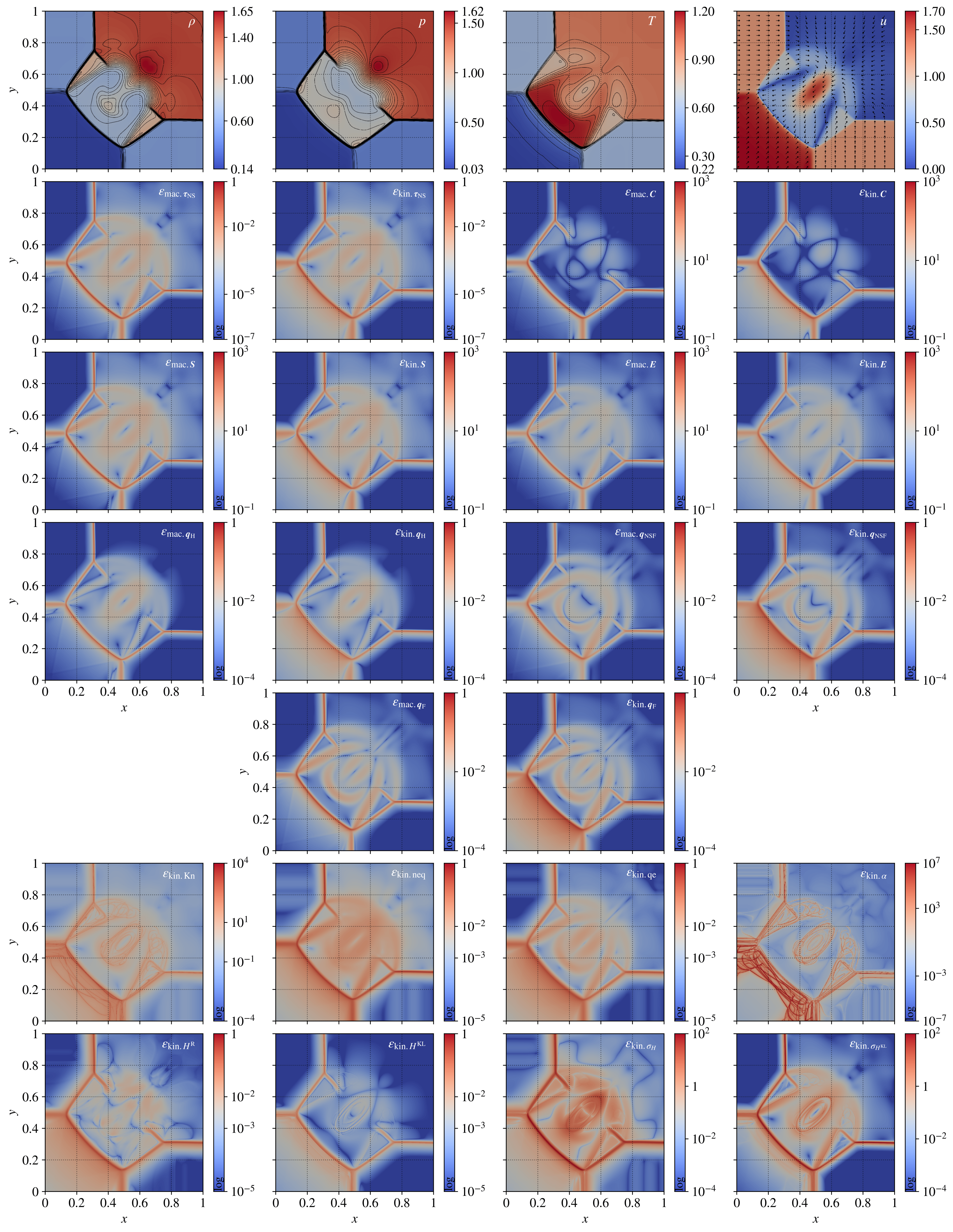}%
    \caption{
    Sensors for the two-dimensional Riemann problem case~No.~$3$ using the 
    internal non-translational energy split
    for a resolution of $\delta x = \delta y = L_x/1024= L_y/1024$.
    Note that the color scale is logarithmic, which is indicated on the respective axis for convenience.
    }
    \label{fig:Val_LR3_Ryk_log}%
\end{figure*}

The resulting sensors 
for different spatial resolutions $\delta x = L_x/128$, $\delta x = L_x/512$ and $\delta x = L_x/2048$ 
are shown in Fig.~\ref{fig:Val_Sod_Tot} for the model with the total energy split,
along the solutions for density, pressure, temperature, flow velocity and respective reference solutions (Riemann solution for the inviscid case) for the sake of comparison.
In addition, individual contributions of the $f$- and $g$-distribution are depicted for some sensors such as the Knudsen, non-equilibrium, quasi-equilibrium and entropic estimate sensors.

A comparison of the individual columns shows the convergence of the sensor values when employing higher resolutions. 
The evaluated sensors generally exhibit greater differences between relatively small and large values and tend to be less noisy and smoother for the same flow features when resolved with finer grids.
In general, the qualitative outcome also has good agreement with what one might expect from intuition based on the reference solutions of the macroscopic variables. This concerns the relative positions and magnitudes of, e.g., flow divergence or momentum stresses, divergence/momentum induced heating, Fourier heat flux and overall energy flux, as well as possible violation of a continuum description, local non-equilibrium contributions, and entropy production.
Some comments and discussions are following for the class 1 and class 2 sensors separately.

(Class 1) 
The kinetic sensors excellently match the macroscopic counterparts computed by gradients of the macroscopic variables, notably, over a range covering several orders of magnitude.
Only slight deviations can be distinguished in the region of the contact discontinuity for the $\epsilon_{\bm{\tau}_{\rm NS}}$ and $\epsilon_{\bm{q}_{\rm H}} = \bm{u}\cdot\bm{\tau}_{\rm NS}$, as well as minor deviations on the positive side of the rarefraction wave for the energy flux sensor  $\epsilon_{\bm{q}_{\rm NSF}}$.
This behavior can be attributed to two observations. 
On the one hand, the macroscopic references of these quantities are computed by central finite-difference approximations, which are susceptible to slight oscillations in the solution of macroscopic variables, and should thus not be interpreted as a ground truth. 
On the other hand, it is expected (from the \ac{CE} expansion) that the macroscopic sensors yield an equivalent expression for the tensors of the order-one expansion in the smallness parameter, which is here compared against the full non-equilibrium tensors utilized in the kinetic computations of these sensors.
This is, as stated in Section~\ref{sec:Refinement-sensors}, only an approximate equality, e.g. $\bm{\tau}_{\rm NS} = 
\bm{P}^{(1)} \approx \bm{P}^{\rm neq} =  \bm{P}^{(1)} + \bm{P}^{(2)} + \dots$, or $\bm{q}_{\rm NSF} = \bm{q}^{(1)} \approx \bm{q}^{\rm neq} =  \bm{q}^{(1)} + \bm{q}^{(2)} + \dots$.
A defect stemming form this approximation is therefore, to some extent, expected to be present and to contribute to the difference in the resulting sensors of the macroscopic versus kinetic computation.
Nevertheless, the precision of the kinetic sensors in class 1 is remarkable, especially considering the sensitive Navier--Stokes--Fourier level heat and energy fluxes.

(Class 2) 
Some insights become visible for the sensors which depict the contributions from the $f$ and $g$ distributions in addition (Knudsen, non-equilibrium, quasi-equilibrium and entropic estimate sensors).
For instance in the quasi-equilibrium sensor, the $f^*$ distribution in the total energy split is per construction equal to the equilibrium, which results in a value of zero, showcasing that the contribution to the QE-sensor stems from $g$ alone.
Moreover, it can be seen that the Knudsen and entropic estimate sensors accommodate very distinct peaks at different locations for $f$ and $g$.
The non-equilibrium sensor reveals that the $g$ distribution is further from $g^{\rm eq}$ than $f$ is from $f^{\rm eq}$, which, in the case of the total energy split, allows to draw direct comparisons between the strength of non-equilibrium with respect to momentum and energy. 
In case such considerations are of relevance in specific applications, the sensors can be applied to either distribution individually.
Moreover, when looking at the entropy related sensors measuring $H^{\rm R}$, $H^{\rm KL}$, $\sigma_{H}$ and $\sigma_{H^{\rm KL}}$, one can see that the versions using the \ac{KL} divergence show a slightly different image and possess a larger variance between relatively small and relatively large values compared to those for $H^{\rm R}$ and $H^{\rm KL}$.
This renders the \ac{KL} versions potentially more useful for the application to \ac{AMAR}.
 
Note that the exact analogue to Fig.~\ref{fig:Val_Sod_Tot} for the model with the internal non-translational energy split can be found in the appendix in Fig.~\ref{fig:Val_Sod_Ryk}.
The same discussion can be concluded for this particular energy split, with the additional comment that, as expected, the outcome differs for some of these kinetic sensors, 
such as the quasi-equilibrium sensor, relative entropy or entropic estimate.
This is due to the differing contributions from the $f$- and $g$-distribution in the compound sensors, stemming from the fundamental difference in the partitioning of the energy onto said distributions in the total versus internal non-translational energy split.

\subsection{Two-dimensional Riemann problem: Case~No.~3\label{sec:val-Riemann-prob}}

The Riemann configurations are further classical benchmarks to investigate the behavior of the solver in two-dimensional compressible flows involving complex interactions between shocks, contact discontinuities, rarefraction waves and vortexes. 
A total of $19$ configurations have been thoroughly studied in the CFD and mathematics literature \cite{LaxLiu1998, KurganovR2D, Schulz-Rinne, ZhangZhengRiemann, TungChang1, TungChang2, RiemannBookZheng, RiemannBookSheng}.
In the following, the modified configuration \#$3$ (case number as introduced by \cite{LaxLiu1998}) is considered, as done by many researchers, e.g., \cite{LR3case, strässle2025a-fully-conservative}.
The modification concerns the borders of the different zones in the initial conditions as well as the simulation time.
The initial conditions read
\begin{align}
    &(\rho, \ p, \ u_x, \ u_y) = 
    \\
    &\left\{
    \begin{aligned}
        &{(1.5 ,\ 1.5 ,\ 0 ,\ 0 ),}    &    &{0.85 \leq x \leq 1, \ 0.85 \leq y \leq 1,}
        \\
        &{(0.5323,\ 0.3,\ 1.206,\ 0),}   &     &{0 \leq x \leq 0.85, \ \ 0.85 \leq y \leq 1,}
        \\
        &{(0.138,\ 0.029,\ 1.206,\ 1.206),}   &     &{0 \leq x \leq 0.85, \ 0 \leq y \leq 0.85,}
        \\
        &{(0.5323,\ 0.3,\ 0,\ 1.206),}    &    &{0.85 \leq x \leq 1, \ 0 \leq y \leq 0.85,}
    \end{aligned}
    \right. \nonumber
\end{align}
in the domain $x \in [0,1]$, $y \in [0,1]$ and the simulation is conducted until $t_{\rm end} = 0.85$s.
A dynamic shear viscosity of \mbox{$\mu = \num{1e-3}$} was imposed and a resolution of  $\delta x = \delta y = L_x/1024 =  L_y/1024$ was adopted.

The resulting sensors using the internal non-translational energy split are depicted in Fig. \ref{fig:Val_LR3_Ryk_log} together with the solutions for density, pressure, and temperature with overlayed contour lines, as well as flow velocity with overlayed Eulerian velocity field indicated by unit arrows, for the sake of comparison.
Note that a logarithmic scale was used for the color scale of the sensors in this figure.
That same results displaying a linear scale can be found in the appendix in Fig. \ref{fig:Val_LR3_Ryk_lin} for reference. 

Overall, the outcome of the resulting sensors is in agreement with what is expected from the knowledge of the macroscopic flow features. 
This two-dimensional case reveals more distinct and clearly separated patterns for each individual sensor, as compared to the previously investigated one-dimensional shock tube configuration.
For the sensors of class 1, an excellent agreement is observed between the kinetic sensors and their macroscopic counterparts computed from gradients of the macroscopic fields, where the same considerations drawn in the evaluation of the previous test case can be confirmed.
\rebuttalUnmark{Within} the sensors of class 2, the Knudsen and the entropic estimate, exhibit very distinct features characterized by relatively small-scale, fine-grained, and continuous structures with elevated values compared to the surrounding. This also translates directly from the observations of the previous evaluation case.
These pronounced and well-resolved patterns in all of the evaluated sensors enable precise and unambiguous identification of targeted flow features of interest.

\section{Application to adaptive mesh refinement\label{sec:ApllicationAMR}}

\begin{figure*}
    \centering
    \includegraphics[height=0.95\textheight, keepaspectratio]{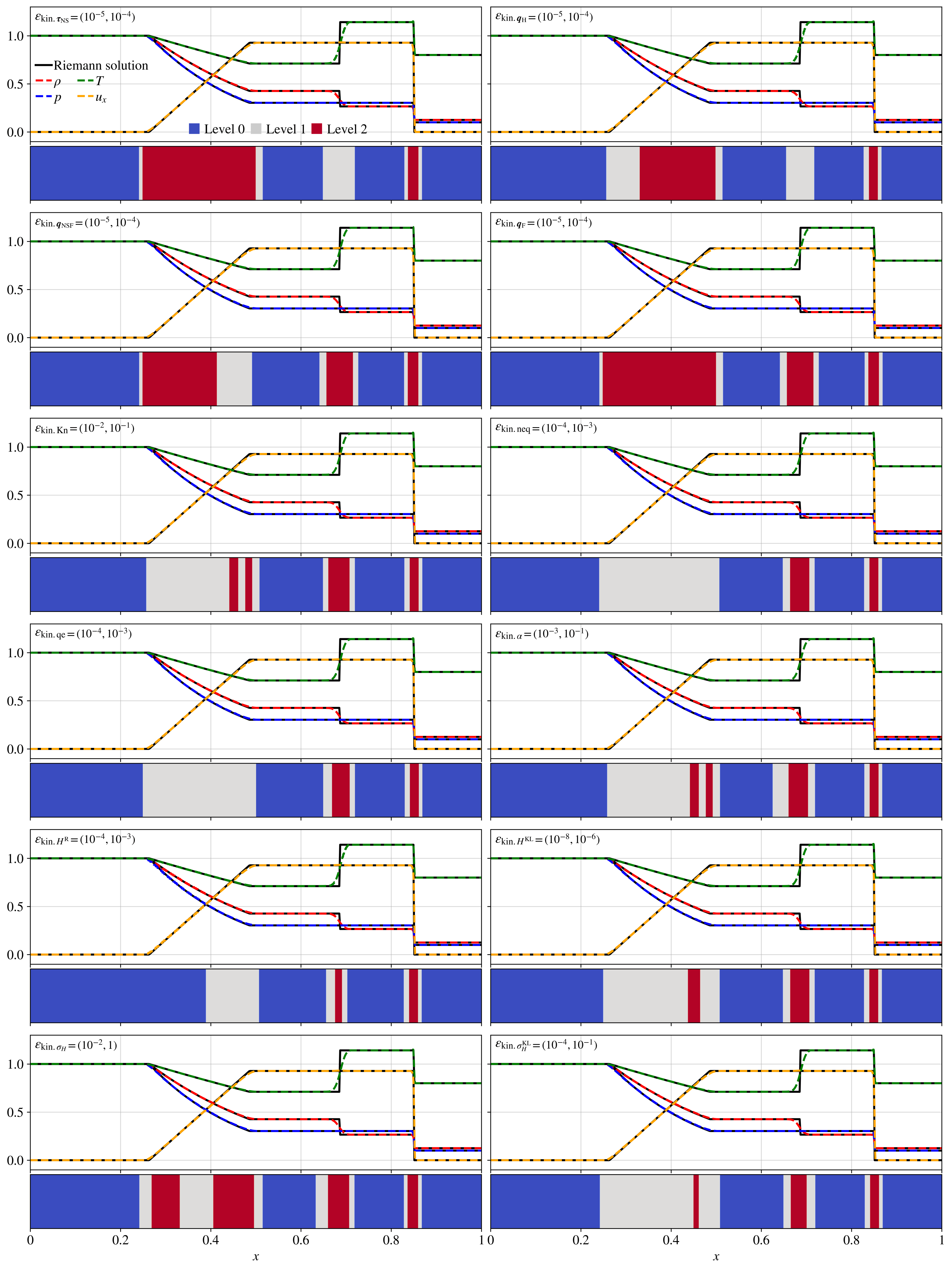}%
    \caption{
    Results and level layout after application of adaptive mesh refinement with the local kinetic sensors for the Sod shock tube 
    using the total energy split.
    The sensor-specific thresholds to trigger the refinement are denoted in each subfigure as $\varepsilon_{\mathrm{kin.}n} = (l_0, l_1)$ for each level, where $l_0$ denotes the threshold on level $l=0$ and $l_1$ denotes the threshold on level $l=1$.
    The levels are indicated by color on the lower axis of each subfigure, where the resolution on the levels are $\delta x = L_x/256$ on level $0$ (blue), $\delta x = L_x/512$ on level $1$ (gray) and $\delta x = L_x/1024$ on level $2$ (red).
    The legends hold for the whole figure. 
    }
    \label{fig:Res_Sod_Tot}%
\end{figure*}

\begin{figure*}
    \centering
    \includegraphics[
    width=0.95\textwidth, 
    keepaspectratio]{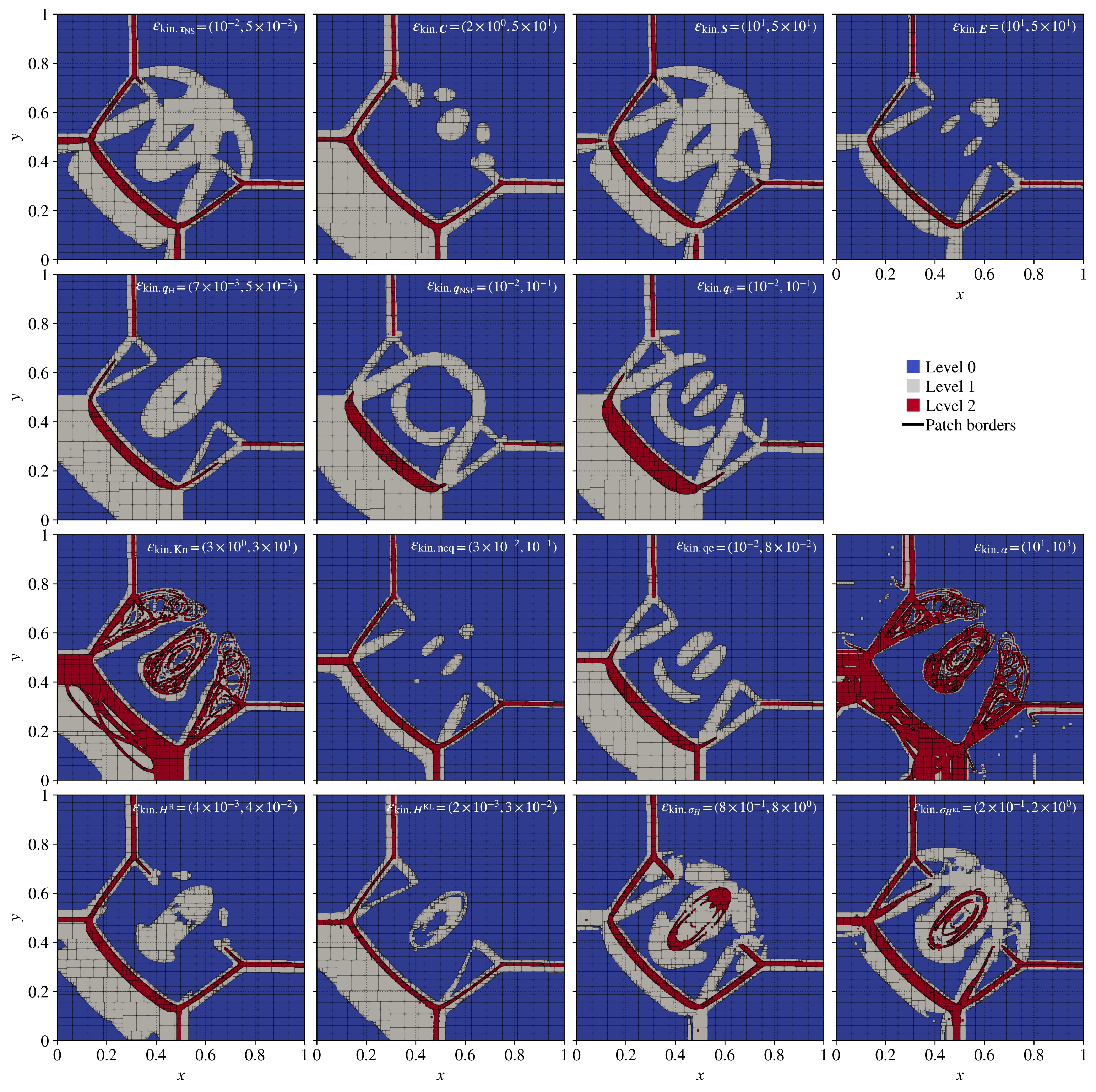}%
    \caption{
    Level layout after application of adaptive mesh refinement with the local kinetic sensors for the two-dimensional Riemann problem case~No.~$3$ 
    using the internal non-translational energy split.
    The level-specific (for level $0$ and level $1$) and sensor-specific thresholds to trigger the refinement are denoted in each subfigure.
    The \ac{AMR} patch borders are indicated by black lines and the levels by color, where the resolution on the levels are $\delta x = \delta y = L_x/512 = L_y/512$ on level $0$ (blue), $\delta x = \delta y = L_x/1024 = L_y/1024$ on level $1$ (gray) and $\delta x = \delta y = L_x/2048 = L_y/2048$ on level $2$ (red).
    The legend holds for the whole figure. 
    }
    \label{fig:Res_LR3_Ryk}%
\end{figure*}
 
In the following the local kinetic sensors are demonstrated with the application of \ac{AMR}.
For the \ac{AMR}, $2$ refinement levels with a constant refinement ratio of $r_l = 2$ across all levels were applied.
Two buffer cells around each patch and a regridding time of $2\delta t_l$ were used for all simulations, as well as a grid efficiency of $70\%$ in one and $98\%$ in two spatial dimensions, respectively.
The same two test cases as used in the validation section were considered, namely the Sod shock tube and the modified Riemann problem case~No.~$3$.
Compared to the validation, thresholds for the selected sensors are imposed on level $0$ and level $1$ to trigger the refinement.
These thresholds were selected and set to refine the main features of the flow such as the shocks or contact discontinuities twice and the secondary features of the flow once.
Otherwise the same considerations, setup and simulation parameters hold as specified in the previous section.

\subsection{Sod shock tube\label{sec:res-Sod-tube}}

The level layouts and results for density, pressure, temperature and flow velocity after application of AMR are depicted in Fig.~\ref{fig:Res_Sod_Tot} for the Sod shock tube, along the reference solutions (Riemann solution for the inviscid case) for the sake of comparison.
The model with the total energy split was applied, and the levels are indicated by color, where the base resolution was selected as $\delta x = L_x/256$ on level $0$, i.e. the resolution on the levels are $\delta x = L_x/256$ on level $0$ (blue), $\delta x = L_x/512$ on level $1$ (gray) and $\delta x = L_x/1024$ on level $2$ (red).

The results demonstrate that the applied \ac{AMR} effectively concentrates resolution in regions of interest, guided by the kinetic sensors. 
By choosing appropriate level- and sensor-specific thresholds, the regions of refinement can be steered. 
This shows that the kinetic sensors can reliably target key flow features and adaptively refine the mesh to capture relevant flow structures while avoiding unnecessary refinement of smooth regions, ensuring computational efficiency and accurate resolution of the critical dynamics.

\subsection{Two-dimensional Riemann problem: Case~No.~3\label{sec:res-Riemann-prob}}

The level layouts of the two-dimensional Riemann problem case~No.~$3$ are depicted in Fig.~\ref{fig:Res_LR3_Ryk} for  the model with the internal non-translational energy split.
The reference solutions can be inferred from Fig. \ref{fig:Val_LR3_Ryk_log}.
The \ac{AMR} patch borders are indicated by black lines 
and the levels are displayed in color, where the base resolution was selected as $\delta x = \delta y = L_x/512 = L_y/512$ on level $0$, i.e. the resolution on the levels are $\delta x = \delta y = L_x/512 = L_y/512$ on level $0$ (blue), $\delta x = \delta y = L_x/1024 = L_y/1024$ on level $1$ (gray) and $\delta x = \delta y = L_x/2048 = L_y/2048$ on level $2$ (red).

It can be seen that the applied \ac{AMR} effectively concentrates resolution in regions of interest guided by the kinetic sensors. 
In this two-dimensional case, the refinement patterns are particularly noteworthy.
Each sensor highlights characteristic flow features in its own distinct way, producing localized and sometimes intricate refinement structures that clearly follow shocks, contact discontinuities, and rarefaction waves on the second refinement level, while highlighting relevant background features of smaller magnitude on level $l=1$. 
\rebuttalUnmark{
It can again be observed that the Knudsen and entropic sensors generate relatively small-scale, fine-grained refinement regions in addition to the regions with discontinuities, which is a characteristic feature of these sensors, cf. Sec.\ref{sec:val-Riemann-prob}.
For the particular choice of thresholds in this 2D Riemann problem, it is interesting to note that the non-equilibrium sensor produced the most compact refinement patterns with very localized refinement along discontinuities, i.e., the smallest refined area for the same accuracy. 
This can be attributed to the fact that this sensor aggregates all dissipative, i.e., non-equilibrium, effects into a single indicator, which can significantly facilitate the selection of thresholds and provide a more general applicability for users. 
}
\rebuttalUnmark{Overall,} the results showcase that, upon choosing appropriate level- and sensor-specific thresholds, the regions of refinement can be steered to reliably target key flow structures and adaptively refine the mesh to capture the most relevant dynamics, while avoiding unnecessary refinement of smooth regions, ensuring both computational efficiency and accurate resolution of critical flow features.

\section{Summary and conclusions\label{sec:Conslusions}}

This work introduced a set of local kinetic refinement sensors designed for adaptive mesh and algorithm refinement within kinetic frameworks, such as discrete velocity and lattice Boltzmann methods. 
While refinement criteria for AMAR based on macroscopic variables can be equivalently replicated from the one-particle distribution function in a purely local, and therefore more scalable, way (class 1), exploiting the accessibility of additional information contained in the distribution function allows to introduce sensors purely accessible in kinetic models (class 2).  
The methodology and implementation were validated against macroscopic counterparts and demonstrated within an adaptive mesh refinement kinetic framework for compressible flows recovering the Navier--Stokes--Fourier equations with variable \rebuttalUnmark{heat capacity} and Prandtl number.

The results confirmed accuracy, robustness and effectiveness of the approach by leveraging distribution-level information in a purely local manner, circumventing the need for non-local gradient computations and thereby improves computational scalability in parallel high-performance computing environments.
Consequently, the framework enhances the fidelity and efficiency of adaptive mesh and algorithm refinement applied to discrete kinetic solvers, establishing a solid foundation for scalable and physically consistent adaptive simulations of complex fluid systems.
Although the sensors are presented for a discrete \rebuttalUnmark{velocity Boltzmann} model for compressible flows with variable \rebuttalUnmark{heat capacity} and Prandtl number, \rebuttalUnmark{as well as the application of adaptive mesh refinement}, these sensors can be used for a wide range of problems, flow scenarios, \rebuttalUnmark{discrete kinetic models and refinement types}.
The ideas can be extended to construct more beneficial sensors for more specific flow scenarios, with some examples provided herein.
\rebuttalUnmark{Furthermore, these ideas can directly be applied with great benefits in the context of adaptive algorithm refinement. 
To name but a few examples among many possibilities, the quasi-equilibrium sensor can be used every $n$-th time step to automatically detect necessary regions where a single-relaxation-time BGK model is insufficient and needs AAR with a more physical but computationally expensive two-relaxation model, 
or the polynomial entropic estimate sensor can be used to determine regions where a more expensive entropic relaxation model, e.g., ELBM or KBC \cite{Karlin_entropicEQ, Karlin_Gibbs, Ali_reviewEntropic}, is required for stability.
}

Future research will extend the proposed refinement sensors toward more specialized flow regimes, including non-ideal fluids and multiphase flows. 
Moreover, the framework offers potential for data-driven optimization of refinement criteria, bridging kinetic theory with modern pattern recognition and machine learning techniques to enable efficient and scalable kinetic simulations across diverse applications in fluid dynamics.


\section*{Acknowledgments}
This work was supported by the European Research Council (ERC) Advanced Grant No. 834763-PonD and by the Swiss National Science Foundation (SNSF) Grant Nos. 200021-228065 and 200021-236715.
Computational resources at the Swiss National \rebuttalUnmark{Supercomputing Centre} (CSCS) were provided under Grant Nos. s1286, sm101 and s1327.
Open access funding was provided by the Swiss Federal Institute of Technology Zürich (ETH Zürich).

\section*{Author declarations}

\noindent\textbf{\textit{Conflict of interest}}\\
The authors have no conflicts to disclose.

\noindent\textbf{\textit{Ethical approval}}\\
The work presented by the authors herein did not require ethics approval or consent to participate.

\noindent\textbf{\textit{AI-assisted technologies}}\\
The authors declare that no AI or AI-assisted technologies were used in the research process, including conceptualization and design of the study, methodology development, data analysis, or interpretation. 
AI-assisted tools were used during the preparation of this manuscript for basic language improvement and writing support.

\noindent\textbf{\textit{Author contributions}}\\
R.M.S.:
Conceptualization of the study, 
formal analysis,
derivation and development of the methodology, 
implementation of the solver, 
evaluation and data analysis, 
writing- initial manuscript and revised versions. 
S.A.H.:
Conceptualization of the study, 
writing- initial manuscript and revised versions. 
I.V.K.:
Conceptualization of the study, 
writing- initial manuscript and revised versions,
funding acquisition, resources.
All authors approved the final manuscript.

\noindent\textbf{\textit{Data availability statement}}\\
The data that support the findings of this study are available within the article or from the corresponding author(s) upon reasonable request.

\noindent\textbf{\textit{Open access}}\\
This article is licensed under a Creative Commons Attribution 4.0 International License, which permits use, sharing, adaptation, distribution and reproduction in any medium or format, as long as you give appropriate credit to the original author(s) and the source, provide a link to the Creative Commons license, and indicate if changes were made. The images or other third party material in this article are included in the article’s Creative Commons license, unless indicated otherwise in a credit line to the material. If material is not included in the article’s Creative Commons license and your intended use is not permitted by statutory regulation or exceeds the permitted use, you will need to obtain permission directly from the copyright holder. To view a copy of this license, visit http://creativecommons.org/licenses/by/4.0/.

\appendix 

\section{Extended evaluation of local kinetic sensors\label{appendix:ExtFigures}}

\begin{figure*}
    \centering
    \includegraphics[height=0.95\textheight, keepaspectratio]{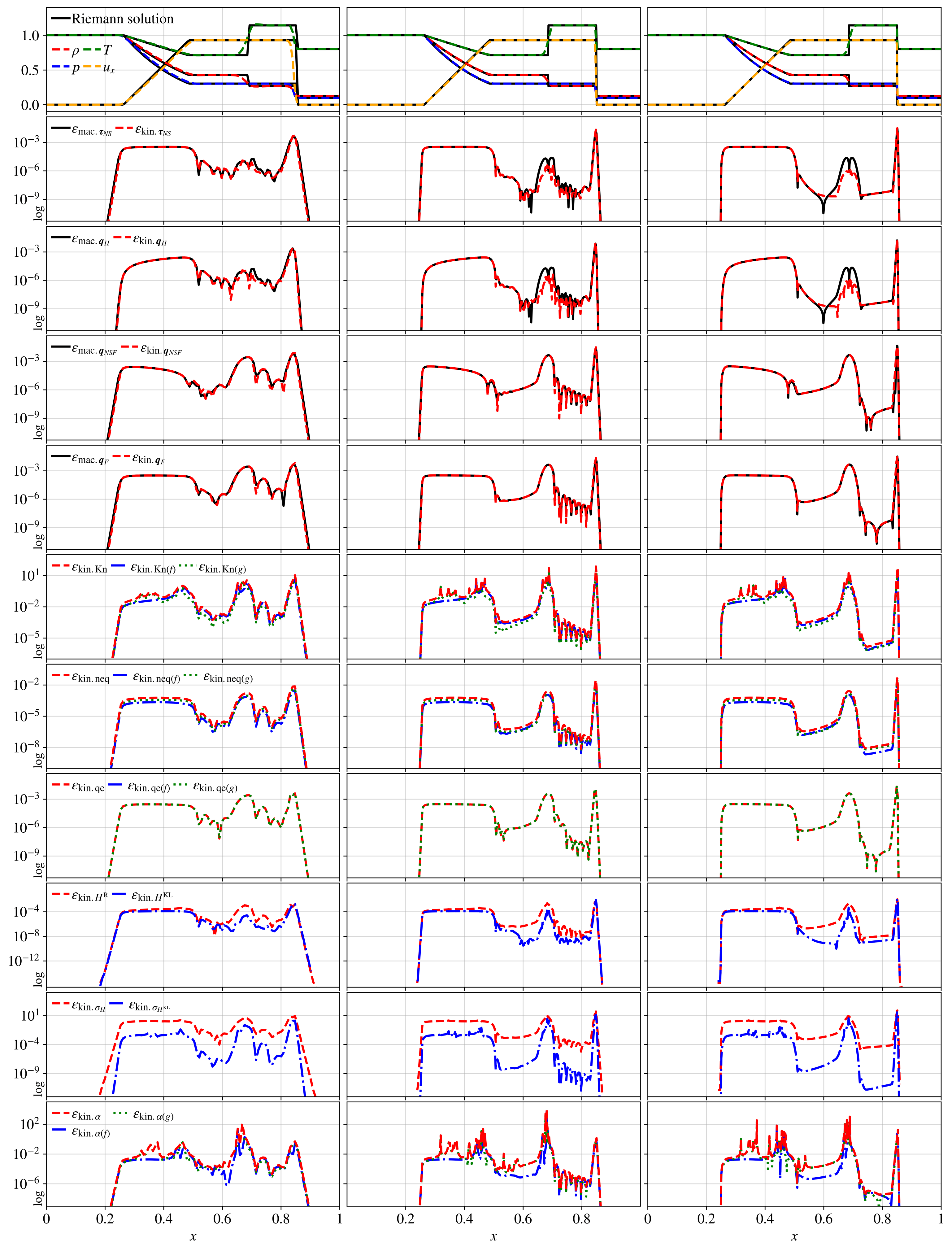}%
    \caption{
    Sensors for the Sod shock tube using the 
    internal non-translational energy split
    for different resolutions depicted from left to right; $\delta x = L_x/128$, $\delta x = L_x/512$ and $\delta x = L_x/2048$.
    The legends hold for the whole row.
    Note that the $y$-axis of the sensors are logarithmic, which is indicated on the respective axis for convenience. 
    }
    \label{fig:Val_Sod_Ryk}%
\end{figure*}

\begin{figure*}
    \centering
    \includegraphics[height=0.95\textheight, keepaspectratio]{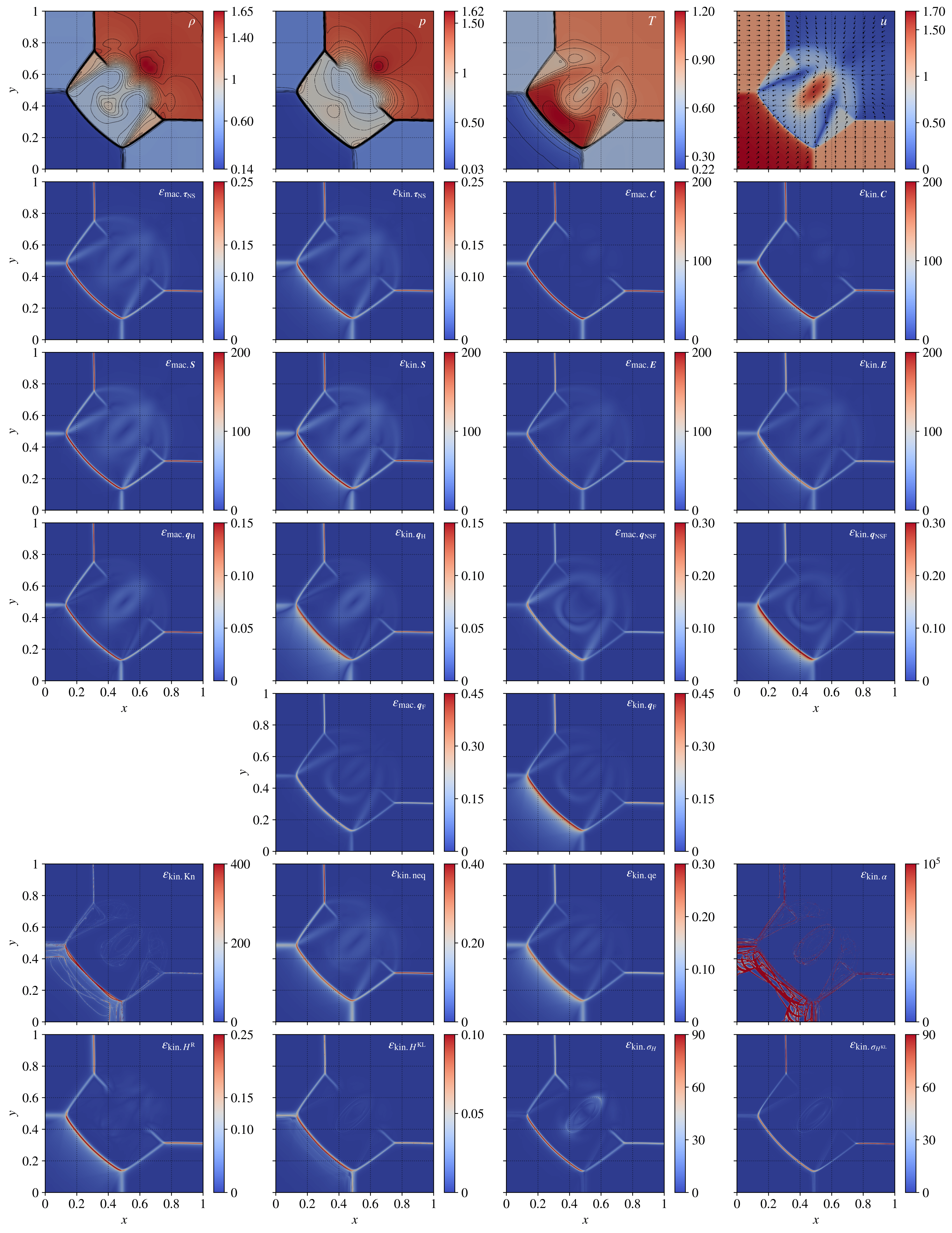}%
    \caption{
    Sensors for the two-dimensional Riemann problem case~No.~$3$ using the 
    internal non-translational energy split
    for a resolution of $\delta x = \delta y = L_x/1024= L_y/1024$.
    Note that the color scale is linear, compared to Fig.~\ref{fig:Val_LR3_Ryk_log}.
    }
    \label{fig:Val_LR3_Ryk_lin}%
\end{figure*}

Fig. \ref{fig:Val_Sod_Ryk} depicts the resulting sensors for the Sod shock tube using the model with the internal non-translational energy split in contrast to to Fig. \ref{fig:Val_Sod_Tot}, where the total energy split was applied.
Different spatial resolutions $\delta x = L_x/128$, $\delta x = L_x/512$ and $\delta x = L_x/2048$ are shown along the solutions for density, pressure, temperature, flow velocity and respective reference solutions (Riemann solution for the inviscid case) for the sake of comparison.
The resulting sensors for the two-dimensional Riemann problem case~No.~$3$ are depicted in Fig. \ref{fig:Val_LR3_Ryk_lin} using a linear color scale in contrast to Fig. \ref{fig:Val_LR3_Ryk_log}, where a logarithmic color scale is depicted.

\clearpage
\section*{References}

%

\end{document}